\documentclass[twoside,twocolumn,9pt]{article}
\usepackage{extsizes}
\usepackage{pdfpages}
\usepackage[super,sort&compress,comma]{natbib} 
\usepackage[version=3]{mhchem}
\usepackage[left=1.5cm, right=1.5cm, top=1.785cm, bottom=2.0cm]{geometry}
\usepackage{balance}
\usepackage{sectsty}
\usepackage{graphicx} 
\usepackage{lastpage}
\usepackage[format=plain,justification=justified,singlelinecheck=false,font={stretch=1.125,small,sf},labelfont=bf,labelsep=space]{caption}
\usepackage{float}
\usepackage{fancyhdr}
\usepackage{fnpos}
\usepackage[english]{babel}
\addto{\captionsenglish}{%
  
}
\usepackage{array}
\usepackage{droidsans}
\usepackage{charter}
\usepackage[T1]{fontenc}
\usepackage[usenames,dvipsnames]{xcolor}
\usepackage{setspace}
\usepackage[compact]{titlesec}
\usepackage{hyperref}

\usepackage{epstopdf}

\usepackage[version=3]{mhchem} 
\usepackage{color}
\usepackage{amssymb}
\usepackage{amsmath}
\usepackage[caption = false]{subfig}
\usepackage[normalem]{ulem}
\usepackage{xr}
\usepackage{booktabs}
\usepackage{multirow}
\usepackage{booktabs} 
\usepackage{comment} 
\usepackage{xr-hyper}

\usepackage{mathptmx}

\definecolor{cream}{RGB}{222,217,201}

\newcommand{\avec}{\textbf{a}}
\newcommand{\bvec}{\textbf{b}}

\newcommand{\Fig}[1]{Fig.~#1} 
\newcommand{\Tab}[1]{Tab.~#1} 
\newcommand{\cm}{cm\textsuperscript{-1}}

\begin{document}


\makeFNbottom
\makeatletter
\renewcommand\LARGE{\@setfontsize\LARGE{15pt}{17}}
\renewcommand\Large{\@setfontsize\Large{12pt}{14}}
\renewcommand\large{\@setfontsize\large{10pt}{12}}
\renewcommand\footnotesize{\@setfontsize\footnotesize{7pt}{10}}
\makeatother

\renewcommand{\thefootnote}{\fnsymbol{footnote}}
\renewcommand\footnoterule{\vspace*{1pt}%
\color{cream}\hrule width 3.5in height 0.4pt \color{black}\vspace*{5pt}} 
\setcounter{secnumdepth}{5}

\makeatletter 
\renewcommand\@biblabel[1]{#1}            
\renewcommand\@makefntext[1]%
{\noindent\makebox[0pt][r]{\@thefnmark\,}#1}
\makeatother 
\renewcommand{\figurename}{\small{Fig.}~}
\sectionfont{\sffamily\Large}
\subsectionfont{\normalsize}
\subsubsectionfont{\bf}
\setstretch{1.125} 
\setlength{\skip\footins}{0.8cm}
\setlength{\footnotesep}{0.25cm}
\setlength{\jot}{10pt}
\titlespacing*{\section}{0pt}{4pt}{4pt}
\titlespacing*{\subsection}{0pt}{15pt}{1pt}

\fancyfoot{}
\fancyfoot[RO]{\footnotesize{\sffamily{1--\pageref{LastPage} ~\textbar  \hspace{2pt}\thepage}}}
\fancyfoot[LE]{\footnotesize{\sffamily{\thepage~\textbar\hspace{3.45cm} 1--\pageref{LastPage}}}}
\fancyhead{}
\renewcommand{\headrulewidth}{0pt} 
\renewcommand{\footrulewidth}{0pt}
\setlength{\arrayrulewidth}{1pt}
\setlength{\columnsep}{6.5mm}
\setlength\bibsep{1pt}

\makeatletter 
\newlength{\figrulesep} 
\setlength{\figrulesep}{0.5\textfloatsep} 

\newcommand{\topfigrule}{\vspace*{-1pt}%
\noindent{\color{cream}\rule[-\figrulesep]{\columnwidth}{1.5pt}} }

\newcommand{\botfigrule}{\vspace*{-2pt}%
\noindent{\color{cream}\rule[\figrulesep]{\columnwidth}{1.5pt}} }

\newcommand{\dblfigrule}{\vspace*{-1pt}%
\noindent{\color{cream}\rule[-\figrulesep]{\textwidth}{1.5pt}} }

\makeatother

\twocolumn[
  \begin{@twocolumnfalse}
\sffamily
\begin{tabular}{m{0.cm} p{17.4cm} }

  & \noindent\LARGE{\textbf{Davydov Splitting Without a Davydov Pair and Highly Mobile Singlet Excitons in Perylene Red Microcrystals}} \\
\vspace{0.3cm} & \vspace{0.3cm} \\

 & \noindent\large{Chris Rehhagen,$^{\dag,\ddag}$\textit{\textsuperscript{\textsection}} Tolibjon Abdurakhmonov,\textit{$^{\dag}$}, Magnus Frank,$^{\dag,\ddag}$, Oliver Kühn\textit{$^{\dag}$} and Stefan Lochbrunner\textit{$^{\dag,\ddag,\ast}$}} \\
\vspace{0.3cm} & \vspace{0.3cm} \\
 & \noindent\normalsize{We investigate the excitonic properties of perylene red microcrystals, whose unit cell contain eight molecules, using both experimental and theoretical methods. Only two of the nominal eight Davydov transitions are experimentally observed, with an apparent splitting of 610~cm\textsuperscript{-1} that is reasonably reproduced by a Frenkel–Holstein exciton model parametrized from density functional theory. A symmetry-based analysis of the eigenstates reveals that the two observed transitions do not belong to the same Davydov pair, so the splitting does not directly report on the Coulomb coupling within the unit cell. The complex mixture of local excitations produces a J-like band for the dominant transitions, favouring exciton transfer through increased spectral overlap. Consistent with this picture, time-correlated single-photon counting of the emission and ultrafast transient absorption spectroscopy show that the dynamics is dominated by highly mobile singlet Frenkel excitons. The exciton mobility extracted from experiment agrees very well with that obtained from Kinetic Monte Carlo simulations, supporting a picture of incoherent hopping transport.} \\

\end{tabular}

 \end{@twocolumnfalse} \vspace{0.6cm}

  ]

\renewcommand*\rmdefault{bch}\normalfont\upshape
\rmfamily
\section*{}
\vspace{-1cm}


\footnotetext{\textit{$^{\dag}$~University of Rostock, Institute of Physics, Albert-Einstein-Str. 23-24, D-18059 Rostock, Germany.}}
\footnotetext{\textit{$^{\ddag}$~Department Life, Light and Matter, University of Rostock, D-18059 Rostock, Germany.}}
\footnotetext{\textit{\textsuperscript{\textsection}~Present Adress: University of Rostock, Institute of Chemistry, Albert-Einstein-Str. 27, D-18059 Rostock, Germany.}}
\footnotetext{\textit{$^\ast$}~corresponding author: stefan.lochbrunner@uni-rostock.de}



\section*{Introduction}
Organic materials are playing an increasingly important role for a wide range of applications such as organic photovoltaics\cite{Wadsworth2019,Hedley2017, Dou2013, Menke2013}, light-emitting diodes,\cite{Salehi2019,Wei2018} field effect transistors \cite{Paterson2018,Luessem2016, Briseno2007} and others\cite{Ostroverkhova2016,Fang2014, Gierschner2016}. 
Besides their cost-efficient production, their key advantages are mechanical flexibility, chemical variability and environment friendliness.\cite{Gressler2022,Seri2021} 
For applications involving the absorption or emission of light, the characteristics and dynamics of excitons are extremely important.\cite{Jiang2021, Park2021, Menke2013, Suman2024} 
Therefore, intense research has been carried out during the last decades to advance our knowledge of the stationary and dynamic properties of molecular excitons.\cite{Young2020,Hestand2018-118,nematiaram21_3368}
Many important findings on the nature of molecular excitons were obtained from dimer systems \cite{Wuerthner2016, Kim2022, Kaufmann2018, Shang2022, Sebastian2021, Mandal2018, Hong2022}, aggregates\cite{Rehhagen2022,Wolter2017,Sung2015,Sorokin2015,Dimitriev2021,Ghosh2026,Shuai2025,Wang2026}, photosynthetic complexes \cite{cao20_eaaz4888,bredas17_35,karki19_7923},  or nanoparticles\cite{Rehhagen2022,Rehhagen2023,Fery-Forgues2013,Pensack2018-122,Jiang2017,Ghosh2026,Wu2006,Tuncel2010} in solution. 
The unknown microscopic structure of aggregates and nanoparticles in solution inhibits a rigorous quantitative comparison of experiment and theory. Therefore, recent  research interest shifted to  extended single crystals.\cite{Brown2024,Volek2024,Fisher2024,Williams2022-144,Myong2021-9,Funke2021,Giavazzi2025-16,Hammer2024}
Crystals offer the  possibility of direct comparison since the arrangement of the molecules can precisely be determined from X-ray data.
Most of the investigated single crystal systems consist of unit cells containing only very few molecules, often one or two\cite{Brown2024,Beljonne2013,Fisher2024}, with a few notable exceptions.\cite{Balzer2022,Muccini1998,Austin2017}
As a consequence, in the vast majority of studied systems no or a simple two-component Davydov-splitting is observed.\cite{davydov_1964,Giavazzi2025-16,nematiaram21_3368}
In fact, under strict  crystal symmetry, four chromophores per unit cell are at least necessary to have three Davydov components (DCs) with a non-vanishing transition dipole moment.
The fourth state is dark due to symmetry.
Actually in the symmetric case no more than three DCs are observable by one-photon processes, no matter of the number of chromophores per unit cell.\cite{Giavazzi2025-162}
Three DCs with non-vanishing transition dipole moment were reported for anilino squaraines crystals with four molecules per unit cell.\cite{Funke2021,Giavazzi2025-16}
Larger unit cells are rarely investigated, especially both experimentally and theoretically and they often show only two DCs.\cite{Austin2017,Muccini1998} \\
In this contribution we investigate crystals of the perylene diimide dye Perylene Red (PR, 1,6,7,12-tetraphenoxy-N,N'-bis(2,6-di-isopropylphenyl)3,4,9,10-perylenedicarboximide), which have an unusually large  unit cell containing eight molecules. Using Frenkel exciton theory parametrized by Density Functional Theory (DFT) calculations, we assign the observed absorption spectrum consisting of an apparent Davydov pair, besides vibronic side bands.
Furthermore, the exciton mobility in terms of diffusive Förster-transfer is studied by time-resolved emission measurements and compared to Kinetic Monte-Carlo (KMC) simulations based on the same Coulomb couplings as the calculated spectra.
In addition, the initial relaxation of the optically generated excitons is characterized by ultrafast transient absorption (TA) spectroscopy. \\
The reminder of the paper is organized as follows. After briefly reviewing the experimental and theoretical methods,
the structural properties of the studied microcrystals are described. 
Then, the polarization dependence of the absorption spectrum is discussed based on experimental and theoretical findings.
Thereafter, the exciton dynamics as measured by time correlated single photon counting is compared to KMC simulations.
Finally, the TA results are described and the conclusions presented. Additional results and analysis are provided in the Electronic Supplementary Information (ESI).

\section*{Methods}

\subsection*{Crystal Preparation}
The sample preparation is based on the evaporation of the solute from a highly concentrated dye solution. The crystals are formed on a glass substrate at the interface of solution and air. This procedure strongly depends on the evaporation rate. Perylene Red (Sigma Aldrich) was dissolved in ethanol at a concentration of 1.4~mM in an ultrasonic bath. The quartz substrate was cleaned with acetone and put upwards but slightly tilted into a glass container filled by 50\% with the PR solution. The containers lip is loose and only fixed to the rest of it with a thin film of paraffin. The container was put into a laboratory hood. The evaporation rate of ethanol can be calculated from the mass difference over time and was adjusted to approximately 10~µm/h.

\subsection*{Stationary Spectroscopy of Single Crystals}
Classical absorption and emission spectroscopy use typically large beam sizes which are unsuited for the spectroscopy of single crystals in the range of tens of micrometers. 
To circumvent this limitation, we build an absorption and emission microscope, see Fig. S1 and S2 in the ESI. 
In short, the sample is illuminated by a focused beam of a halogen lamp and a 20x magnified image of the crystal is generated with an objective. 
The image is split by a beam splitter.
 50\% is lead onto a CMOS camera and the rest is guided to a spectrometer (QE65000, Ocean Optics) via a fibre with a diameter of 50~µm.
The fibre picks out a small area of the magnified image. 
This area corresponds to a size of 2.5~µm in the plane of the sample, i.e. 50~µm divided by the magnification of 20, which is the spatial resolution. 
By comparing the transmission with and without crystal its absorption can be determined. The polarization of the white light is adjusted with a wire-grid polariser. \\
For emission microscopy, the same set-up is used but with 532~nm excitation instead of a halogen lamp. 
The excitation can be done both in transmission and in reflection geometry, i.e. either through the objective or from the direction of the white light.

\subsection*{Confocal Microscopy}
The thickness of the sample is determined by a confocal microscope (OLS4100, Olympus) with 50x optical magnification. The error of the measured thickness is about 100$~$nm.

\subsection*{X-Ray Diffraction}
A X-ray quality PR crystal was selected in Fomblin YR-1800 perfluoroether (Alfa Aesar) at ambient temperature. The sample was cooled to 123(2) K during the measurement. The data was collected on a Bruker D8 Quest diffractometer using Mo $K_\alpha$ radiation ($\lambda = 0.71073$~\AA). The structure was solved by iterative methods (SHELXT)\cite{Sheldrick2015a} and refined by full-matrix least squares procedures (SHELXL).\cite{Sheldrick2015b} A semi-empirical absorption correction was applied (SADABS).\cite{Sheldrick2016} All non hydrogen atoms were refined anisotropically, hydrogen atoms were included in the refinement at calculated positions using a riding model.

\subsection*{Fluorescence Lifetime Imaging Microscopy}
Spatially resolved emission dynamics is detected with a fluorescence lifetime microscope (MT200, PicoQuant). The sample is excited with a 120~ps (FWHM) laser pulse at a centre wavelength of 442~nm with a repetition rate of 40~MHz. An objective with 20x magnification is used for both focusing the excitation beam and collection of the emitted fluorescence. The excitation spot size is 1.6~µm (FWHM) and the probed area 1.2$~$µm (FWHM). The fluence is varied between 0.04~µJ/cm$^2$ and 90~µJ/cm$^2$, ranging over more than three orders of magnitude.

\subsection*{Femtosecond Pump-Probe Microscopy}
The setup for the pump-probe microscopy is schematically shown in Fig. S3 in the ESI.
It is based on a regenerative Ti:sapphire laser system(Clark MXR , CPA 2001) providing pulses centred at 775~nm with a duration of 180~fs at a repetition rate of 1~kHz. The output is split into two beampaths for the generation of pump and probe pulses.
As probe pulse serves a white light supercontinuum generated in a 6~mm thick sapphire crystal pumped with a small fraction of the fundamental. 
It is focused by an off-axis parabolic mirror with a focal length of 50.8$~$mm to a focus size of 7~µm (FWHM).
The pump pulses centred at 540~nm are generated with a noncollinear optical parametric amplifier and compressed with a fused silica prism compressor to a pulse length of 30~fs.
The beam is focused by a spherical mirror with a focal length of 500~mm to a focal size of 200~µm.
The pulse intensity is adjusted by two subsequent wire-grid polarisers.
The polarizations of pump and probe are adjusted by an achromatic and a superachromatic half-wave plate, respectively (RAC 3.2.10 and RSU 1.2.10, Bernhard Halle Nachfl.).
The time delay between the pump and the probe pulse is adjusted by a retro reflector on a motorized delay stage in the probe beam path before the supercontinuum generation.
A chopper in the pump beam path blocks every second pump pulse such that two subsequent white light pulses probe the excited and the unexcited sample and the absorption change is calculated from the ratio of these two signals.
The sample is placed in the focal plane of the beams.
An objective (20x, NA 0.4) generates a magnified image of the crystal and the probe beam spot.
A fibre with a core diameter of 100 µm is placed in the image plane. 
It collects only the light within the area given by its core, corresponding to a spatial resolution of 5 µm in the plane of the crystal. 
The fiber guides the light to a home-built grating spectrometer. 
In it, the beam is spectrally dispersed with a grating (GR50-0605, Thorlabs) and focused onto a CCD line detector (Entwicklungsbuero Stresing). 
The spectral resolution is 2 nm and the spatial resolution 5 µm.
The time resolution is 200~fs, as extracted from the signal rise time.

For alignment purposes, a beam splitter can be placed in the beam path after the objective.
The reflected part of the beam is guided onto a CMOS camera.
There, the overlap of pump and probe on the crystal can be adjusted.
Besides, a thin glass substrate can be placed in the probe beam path in front of the off-axis parabolic mirror.
It couples the light of a halogen lamp into the probe beam path.
The focus size of the halogen lamp light is large enough to illuminate the entire crystal and its surrounding.
As a consequence, a complete microscopy image of the crystal with both pump and probe visible can be recorded.

\subsection*{Hamiltonian and Parametrization}
Absorption spectra were calculated employing the  Frenkel-Holstein exciton model.\cite{may23,schroter15_1a,hestand18_7069}  Hamiltonian parameters, i.e.  transition dipole moments and Coulomb couplings were obtained by linear response time-dependent density functional theory (TDDFT) using the crystal structure determined by X-ray diffraction (Fig.~\ref{fig:PR_microscopy_unitcell}d). To calculate the excitonic couplings efficiently,
the transition densities were approximated by  transition point charges. Periodic boundary conditions were applied to consider the contributions from interactions between equivalent and non-equivalent molecules in the crystal. The vibronic coupling was modelled by a single vibrational mode per molecule with vibrational frequency $\hbar \omega = 1420$ \cm and a Huang-Rhys factor $S  = 0.64$, estimated according to the experimental data. The vertical electronic excitation energies are considered to be identical for all monomers and taken from the experimental monomer spectrum, $\varepsilon_m=17316.9$ \cm. Further details can be found in Section S6, ESI.

\subsection*{Kinetic Monte Carlo}
The incoherent exciton diffusion is modelled as a Markov process using the  event-driven Kinetic Monte Carlo (KMC) method. At each instant we consider   three stochastic events for an exciton, i.e. its intrinsic decay, exciton hopping from one site to another, and exciton-exciton annihilation. Every event is assigned to a rate constant that reflects its underlying physics. The intrinsic decay rate, i.e. the inverse of the exciton lifetime is taken from the FLIM measurements. Annihilation is is assumed to proceed instantaneously once two excitons meet at the same site. For the exciton diffusion a F\"orster  transfer of  localized excitations is assumed.  Thus the donor-acceptor transfer rate $k_{\rm DA}$ is calculated via F\"{o}rster theory:\cite{Foerster1949,may23}
\begin{equation}
    k_{\rm DA} = \frac{2\pi}{\hbar}V_{\rm DA}^2 \,J,
    \label{eq:forster_rate}
\end{equation}
where $V_{\rm DA}$ is the electronic coupling between donor and acceptor sites and $J$ is the spectral overlap. The electronic coupling is taken from the results of the TDDFT calculations performed for simulating the absorption spectra and described in Section S6 and the spectral overlap from the experiment, see Section S8 in the ESI for details. \\
The KMC simulation was run in a supercell of $[30\times 30 \times 30]$ unit cells taking into account periodic boundary conditions (PBC). Two initial exciton densities were considered, corresponding to the two highest excitation levels applied in the experiments. Only the highest densities are considered since for lower densities an increased supercell size would be required to get accurate statistics which is not affordable concerning computational cost. For each exciton density 64 individual trajectories were sampled and averaged to calculate the time evolution of the exciton density.

\section*{Results and Discussion}
\begin{figure}[t]
\centering
  \includegraphics[width = 1\linewidth]{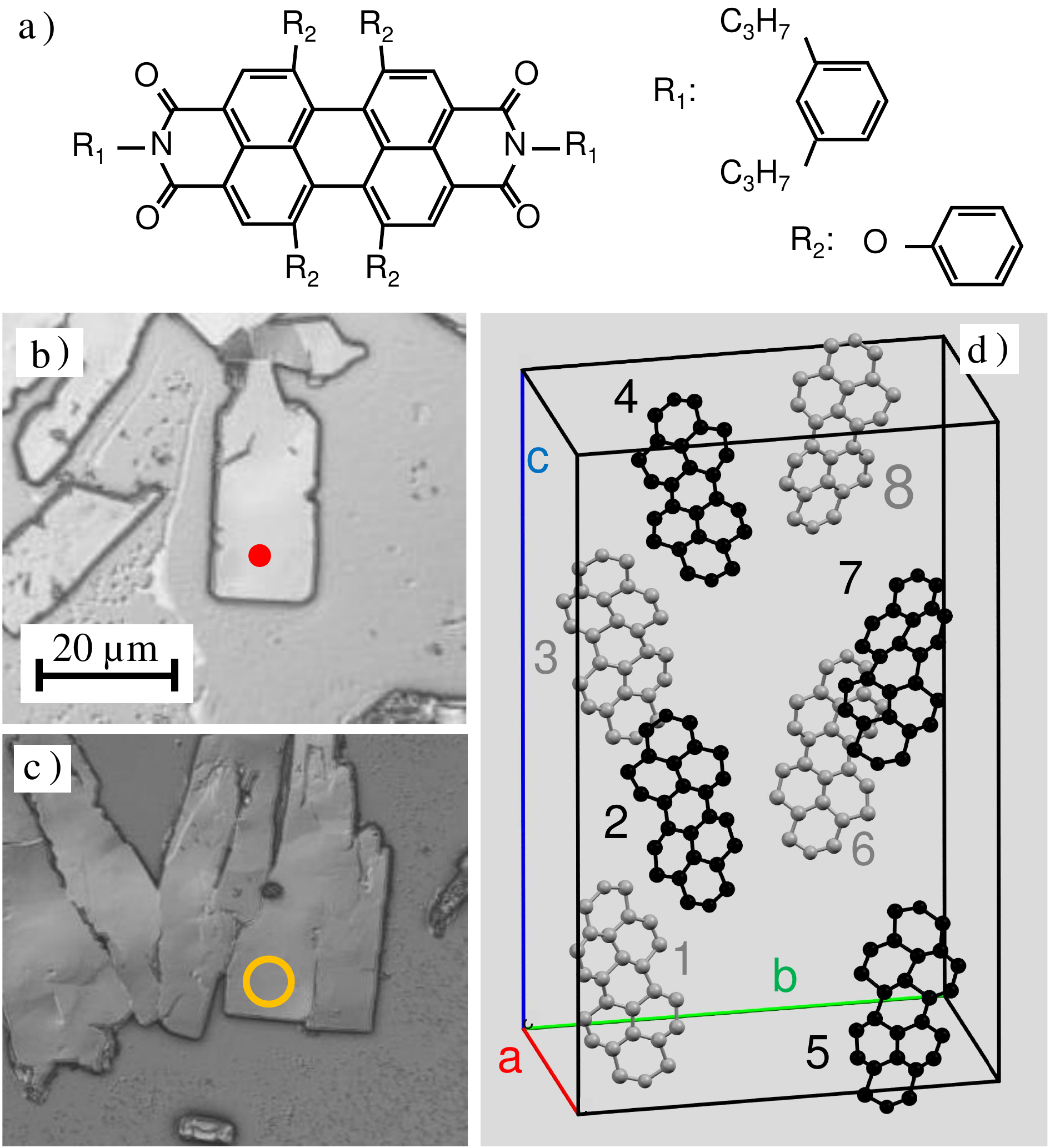}
  \caption{Molecular structure of PR and images of typical crystals and their unit cell. (a) Molecular structure of the PR molecules. (b and c) Confocal microscopy images of two typical crystals, called crystal A and D in the following, see section S2 in the ESI for an overview of the investigated crystals. The red dot shows the spatial resolution of the stationary absorption measurements and the orange ring the spatial resolution of the TA experiments. 
 (d) Unit cell of the crystal structure. For simplicity, the substituents at the bay and the imide positions and the hydrogen atoms are not shown. The cell contains eight molecules and has a size of 1.35~x~2.31~x~3.82~nm$^3$ along the crystal axis {\bf a} (red), {\bf b} (green) and {\bf c} (blue). More information on the arrangement in the unit cell can be found in section S2 in the ESI.}
  \label{fig:PR_microscopy_unitcell}
\end{figure}
\subsection*{Shape and Structure of the Crystals}
Single crystals of the dye Perylene Red (PR) were prepared as described above.
Confocal microscope images of investigated crystals are shown in \Fig{\ref{fig:PR_microscopy_unitcell}}. They are flat and 1.4 to 2.5~µm thick. Typically, the crystals have a rectangular shape with a short and a long edge. \\
X-ray diffraction on the crystals shows that the molecules crystallize in a crystal structure with Pca2$_1$ space group symmetry and orthorhombic unit cells. Those consist of eight molecules and have a size of 1.35~x~2.31~x~3.82~nm$^3$, see \Fig{\ref{fig:PR_microscopy_unitcell}}. The average volume per molecule is 1.49~nm$^3$.
The surprisingly large number of molecules in the unit cell results from the repetition of two slipped-stacked molecules with small deviations in their relative orientation as well as in the angle of two neighbouring pairs. As a consequence, the molecules appear in a wave-like arrangement along the \bvec-axis and in a trapezoidal formation perpendicular to the \avec-axis, see section S2 in the ESI.

\subsection*{Absorption Spectra of Single Crystals}
\begin{figure}[t]
\centering
  \includegraphics[width = 1\linewidth]{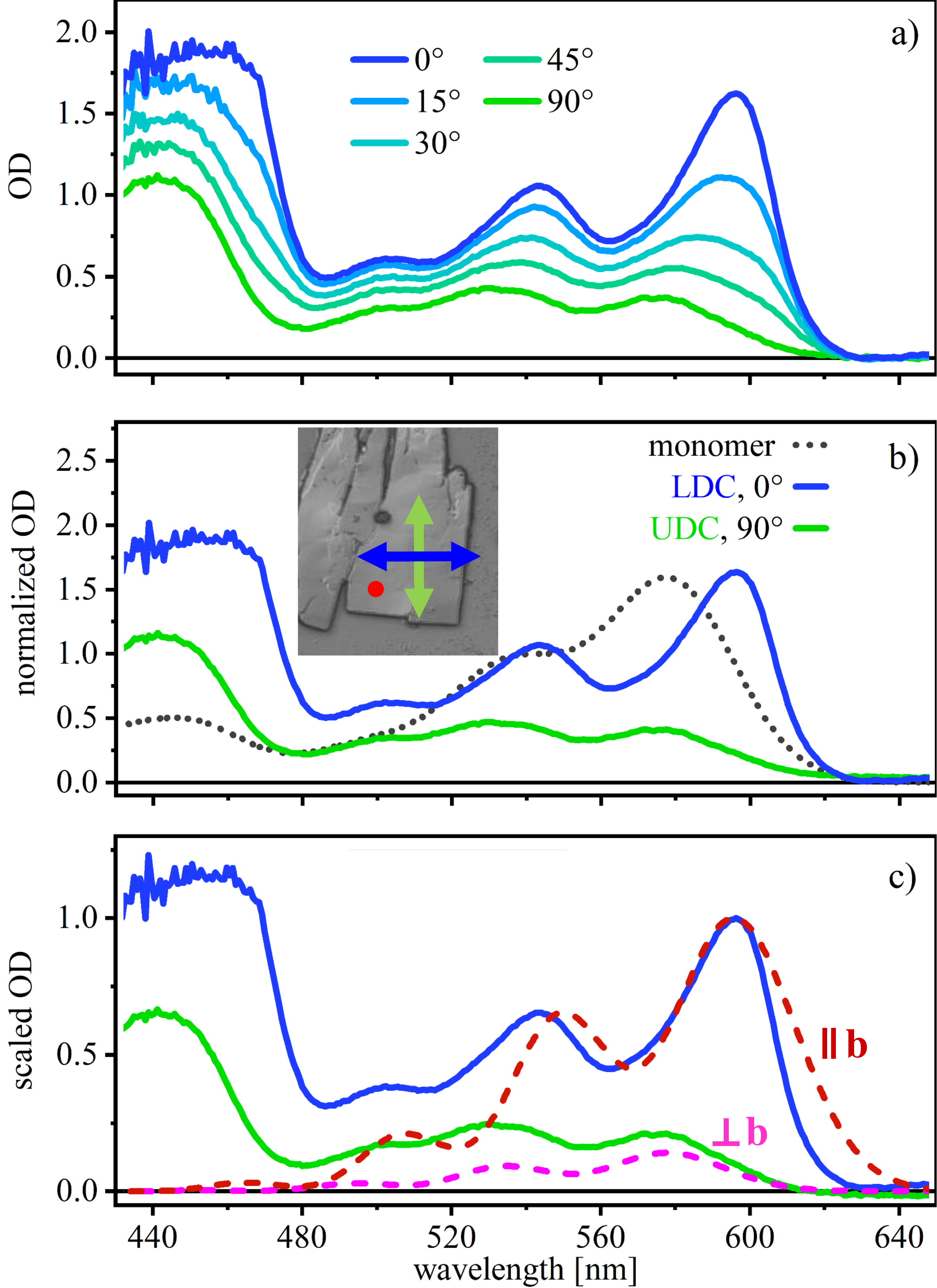}
  \caption{Absorption spectra of crystal D shown in the inset of panel b). The red dot indicates the point of measurement. The arrows indicate the polarization. Panel a): Polarization dependence of the absorption. In the legend, the relative angles between the short edge of the crystal and the polarization are given. An OD above 1.7 is not measurable. Therefore, the spectral shape is flattened below 480~nm. Panel b): Spectra of the two apparent Davydov components and that of the monomer in ethanol. The 0°-absorption is called LDC absorption and the 90°-absorption UDC absorption in the following. Panel c): Comparison of experimental (solid) and calculated (dashed) spectra. For the calculated spectra the transition dipole orientation relative to the \bvec-axis of the unit cell is given, see \Fig{\ref{fig:PR_microscopy_unitcell}}.}
  \label{fig:abs}
\end{figure}

In \Fig{\ref{fig:abs}} the polarization dependent stationary absorption spectrum of a single crystal is shown together with its decomposition in two components polarized parallel and perpendicular to the {\bf b}-axis.
Similar to the monomer in solution, the crystal spectra exhibit two bands with maxima at about 600~nm and 440~nm which are related to the S$_0$-S$_1$ and S$_0$-S$_2$ transition in the monomer. The former band exhibits a vibrational progression with two further maxima between 500 and 560~nm. \\
The measured optical density of the single crystal spectrum in the maximum at 600~nm is 1.7. This is far below what is expected for an extinction coefficient of about 45,000~Mcm$^{-1}$ found in solution,\cite{Colby2010} a concentration in the crystal of $\approx$1~M and a thickness of a few µm. We attribute this feature to an upright orientation of the molecules relative to  the substrate as the transition dipole of the S$_0$-S$_1$ transition is aligned along the long molecular axis (see Fig. \ref{fig:state_rep} below), which is mostly located in the {\bf b}-{\bf c}-plane, see \Fig{\ref{fig:PR_microscopy_unitcell}}d. Since the {\bf c}-axis is perpendicular to the crystal surface and thereby also to the polarization of the incident light, only the small projection of the transition dipole on the {\bf b}-axis can interact with the light field. 
A further result of this orientation of the unit cell is the relative strength of the S$_0$-S$_2$ transition (440~nm) of the crystal which is much larger than the S$_0$-S$_1$ one, contrary to the monomer. This results as the transition dipole of the S$_0$-S$_2$ transition is along the short molecular axis and thereby parallel to the crystal surface and in the plane of the light polarization. \\
The absorption band at 600 nm and its vibronic progression exhibit a pronounced polarization dependence. The absorption is maximal if the light is polarized parallel to the short edges of the crystals. We define the polarization angle therefore relative to this axis. 
Changing the angle from 0° to 90°, i.e. rotating the polarization from parallel to perpendicular to the short crystal edges, leads to an almost 4-fold reduction of the absorption strength and a spectral shift of the maximum from 596~nm a 0° to 575~nm at 90°, corresponding to an energy difference of 610~cm$^{-1}$. 
Accordingly, from the spectra two perpendicular polarized components can be reconstructed which are shifted by 610~cm$^{-1}$.
Typically this is interpreted as Davydov splitting \cite{davydov_1964} and the observed transitions are called lower (LDC) and upper Davydov component (UDC) in the following. Whether the splitting in the present case  corresponds indeed to a simple Davydov pair will be discussed below. \\
Figure \ref{fig:abs}c) compares measured and calculated spectra for the two polarization directions. The peak positions and the intensities of the 0-0-transition as well as the  vibronic progression are well reproduced thus validating the present model. We also find a reasonably good agreement between the experimentally observed splitting of 610~cm$^{-1}$ with the calculated value of 520~cm$^{-1}$. \\
Besides the polarization dependence of the overall amplitude and the position of the maxima, the vibrational progression is polarization dependent as predicted by Spano and coworkers.\cite{Giavazzi2025-162} 
The UDC absorption shows an H-like vibrational progression whereas it is J-like for the LDC absorption.
This is visible by the increased amplitude ratio A$_1$/A$_2$ of the 1\textsuperscript{st} and the 2\textsuperscript{nd} maximum of the vibrational progression for the LDC absorption and vice versa for the UDC absorption.\cite{Giavazzi2025-162} \\
\begin{figure*}[t]
 \centering
 \includegraphics[width = 0.7\linewidth]{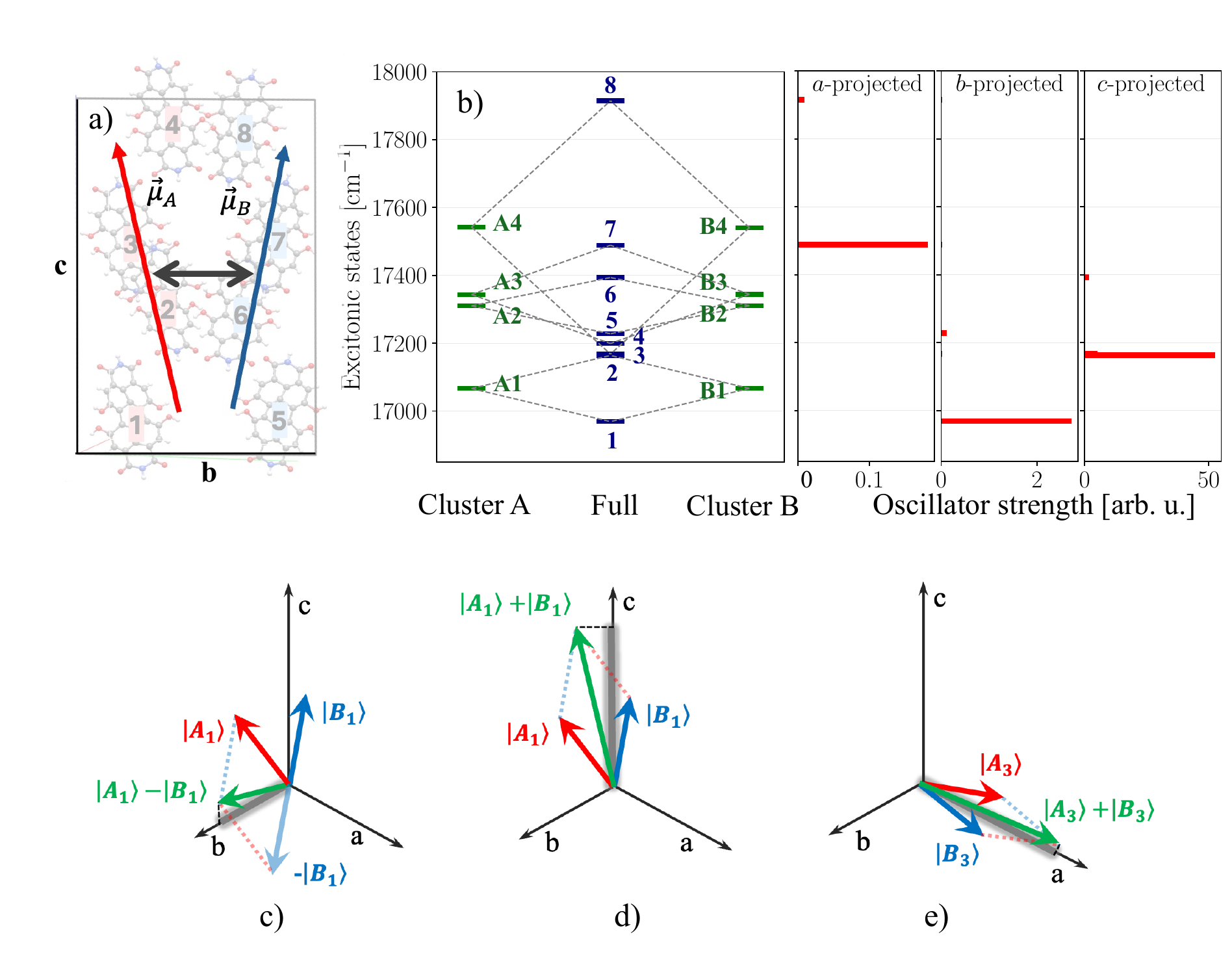}
 \caption{Correlation diagram and the representation of the unit cell. (a) Arrangements of two tetramer clusters A and B in the unit cell and their cluster transition dipole moments, $\vec{\mu}_A$ and $\vec{\mu}_B$ (sketch). (b) Correlation diagram showing how the local cluster exciton states (A1-A4, B1-B4) combine into full eigenstates states (1-8). Dotted lines highlight only the highest overlaps above a threshold value of 0.25. The panels on the right show the oscillator strength of each of the eight exciton states projected onto the crystallographic \textbf{a}-, \textbf{b}- and \textbf{c}-axes. The measured apparent  Davydov splitting is identified as the energy difference between the lowest and the highest state with sufficient oscillator strength, i.e. states 1 and 7 in this case. (c-e)  Schematic vector representation of the coupled cluster states for $k=1, 2$ and 7 showing the origin of the bright states and their polarization.}
 \label{fig:state_rep}
\end{figure*}
For a dimeric unit cell the Davydov splitting is expected to be twice the Coulomb coupling between the monomers. The calculated maximum Coulomb coupling is about 200 \cm, i.e. not sufficient to give a  Davydov splitting of 520~cm$^{-1}$. In any case the unit cell contains eight coupled monomers.
In order to rationalize the appearance of two absorption peaks  an in-depth analysis of the Hamiltonian matrix and its eigenstates is provided in the ESI, Section S7; to ease the argument  we neglect vibronic coupling. Central to the analysis is the observation that the eight monomers in the unit cell can be grouped into two clusters, i.e. A (1-4) and B (5-8) in Fig. \ref{fig:PR_microscopy_unitcell}d). These two clusters have nearly exchange symmetry, i.e. upon swapping indices of the Hamiltonian matrix one obtains a nearly identical matrix. This suggests to calculate the eigenstates of the Hamiltonian in block-diagonal approximation, which will be called cluster eigenstates. Interestingly, the residual coupling between the cluster eigenstates is large only for those pairs of states that are nearly resonant. The couplings become as large as 370 \cm.
The emerging picture is sketched in Fig. \ref{fig:state_rep}a). The collective cluster eigenstates mix in $\pm$ combinations to give rise to Davydov-split pairs of states. From the quantitative analysis of the full eigenstates in terms of cluster states and the respective transition dipole matrix elements, the absorption spectrum can be explained in terms of the correlation diagram provided in panel b) of Fig. \ref{fig:state_rep}.
Labelling the cluster states as $A_i/B_i$ ($i=1\ldots 4$) the eigenstates of the full Hamiltonian, labelled $k=1 \ldots 8$, contribute as follows to the absorption spectrum.  For {\bf c} axis projection, the strongest transition is into the full eigenstate $k=2$ which is the plus combination, $A_1+B_1$, of cluster eigenstates. The respective minus combination, $A_1-B_1$, i.e. the Davydov pair state,  corresponds to the full eigenstate $k=1$. It is polarized along the {\bf b} axis.  Along the {\bf a} axis it is  the plus state $A_3 + B_3$ ($k=7$) that carries oscillator strength. The related $A_3 - B_3$ ($k=4$) combination has negligible oscillator strength.  The direction dependent spectra are shown in Fig. \ref{fig:state_rep}b). The vector addition of the transition dipoles of cluster eigenstates is shown in panels c)-e). Thus we can conclude that the observed difference between peaks in {\bf a} and {\bf b} polarization (calculated at 520 \cm) is not related to the energy difference of a $\pm$ split pair of local cluster states. Instead, these peaks belong to two \emph{different} Davydov pair states. Hence it is not surprising that the apparent splitting does not correspond to any Coulomb coupling matrix elements. This implies that the splitting cannot be used to estimate the Coulomb coupling strength as well. \\
Finally, we note that the calculated oscillator strengths are rather strongly depend on the crystal axis projection. The transition carrying the by far highest oscillator strength is {\bf c} polarized. It is not detected in the experiment which measures the projection onto the {\bf a}-{\bf b}-plane. However, the much lower oscillator strength of the optically accessible directions, i.e. transitions to $k=1, 7$, is inline with the rather low optical density of the crystal discussed in the context of \Fig{\ref{fig:abs}}. \\
In contrast to this finding, for a comparable system with eight molecules per unit cell the appearance of only two DCs was explained by the dominance of the nearest neighbour interaction.\cite{Austin2017} 
Indeed, comparing the unit cell geometries between the present crystals and those reported in Ref. \cite{Austin2017} one clearly notices that in the latter case the corresponding A and B clusters dimerize, such that the absorption is dominated by the strongly coupled $\pi$ stacked dimers, which is in contrast to the present crystal featuring more uniformly distributed couplings.

\subsection*{Exciton Dynamics at Low Excitation Powers}
\label{section:results_FLIM}
\begin{figure}[t]
\centering
  \includegraphics[width=1\linewidth]{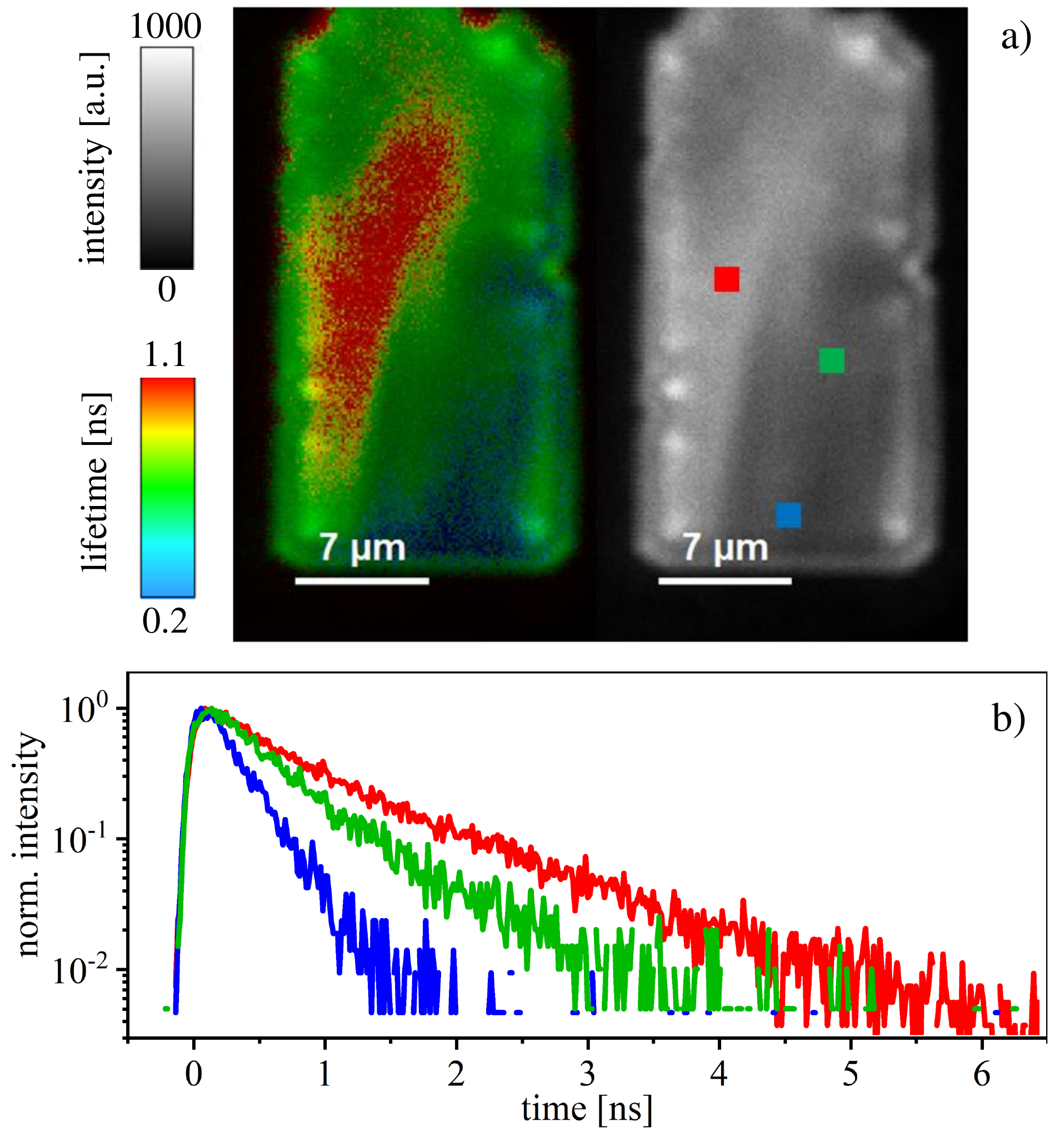}
  \caption{FLIM results of PR crystal A. Upper panel: Spatially resolved lifetime. The color indicates the lifetime, and the brightness correlates with the emission intensity. On the left side, the image with both lifetime and intensity is shown, in the right part of the image, only the intensity distribution. Lower panel: Time traces of the emission after excitation at 442~nm at low photon flux. The data was taken at three different positions on the crystal as indicated by squares of the same colour in the upper right image. The three areas corresponds to area 1 (red), 2 (green) and 3 (blue) in \Tab{\ref{tab:FLIM_results}}}
  \label{fig:FLIM}
\end{figure}

We analyse the exciton dynamics by spatially resolved fluorescence lifetime imaging microscopy (FLIM)\cite{Datta2020} as the fluorescence intensity is a measure for the exciton density. In FLIM, the time-dependence of the emission is detected via time-correlated single photon counting (TCSPC) with a spatial resolution of $\approx$ 1~µm; see above for further details.
At a low excitation flux, the collective fluorescence of the entire crystal exhibits a multi-exponential decay but shows exponential behaviour if only small areas of the crystals are considered, see \Fig{\ref{fig:FLIM}}. The lifetimes vary between 300 and 1000~ps within the same crystal. The lifetime correlates with the emission intensity but the emission strength at time zero is identical over the entire crystal. We conclude that the emission quantum yield is spatially inhomogeneous. In addition the decay rate is dominated by a non-radiative relaxation channel since the monomer lifetime of PR is about 6~ns.\cite{Rehhagen2022,Al-Kaysi2006}
We hypothesize that the non-radiative decay channel is related to defects in the crystal and the local defect density determines the rate of the exciton decay.

\subsection*{Exciton Interaction and Diffusion}
\begin{figure}[h]
\centering
  \includegraphics[width=1.0\columnwidth]{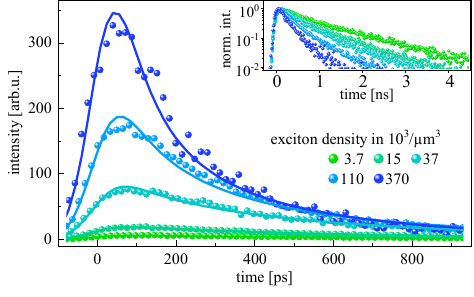}
  \caption{Emission time traces of the crystal A with increasing exciton density. The solid lines show the fit results. In the inset, time traces normalized to the peak intensity are shown.}
  \label{fig:FLIM_timetrace}
\end{figure}

\begin{table}[b]
\small
  \caption{Fit results for the annihilation constant $\gamma$ and the diffusion constant $D$ for different crystals and different areas in there. Crystals A-C are shown S2 of the ESI. Crystal A is also shown in panel b) of \Fig{\ref{fig:PR_microscopy_unitcell}}. The given lifetimes are prior to photodegradation, see section S4 in the ESI for more details.}
  \label{tab:FLIM_results}
  \begin{tabular*}{0.48\textwidth}{@{\extracolsep{\fill}}ccccccc}
    \hline
    & \multicolumn{6}{c}{crystal \& area} \\
    \cmidrule(lr){2-7}
    & A & A & A & B & B & C \\
    & area 1 & area 2 & area 3 & area 1 & area 2 & area 1 \\
    \hline
    $\tau$ [ps] & 1000 & 600 & 350 & 550 & 750 & 800 \\
    $\rm\gamma$ [$\rm nm^3/ps$]& 1.2 & 1.1 & 1.3 & 1.5 & 1.9 & 2.6 \\
    D [$\rm nm^2/ps$]& 2.9 & 2.6 & 3.1 & 3.6 & 4.5 & 6.3 \\
    \hline
  \end{tabular*}
\end{table}
The time-resolved emission measurements at different excitation powers show that with increasing exciton density the fluorescence decay accelerates, see \Fig{\ref{fig:FLIM_timetrace}}. 
This is a typical signature of exciton-exciton annihilation (EEA). EEA describes the process where two excitons interact to yield one monomer in a higher electronically excited state and the other one in the ground state. The higher electronically excited state rapidly deactivates to the S$_1$ state due to internal conversion. As a result the electronic excitation energy of one exciton is converted into heat (vibrational motion). For the fluorescence decay this leads to an acceleration  depending on the density and mobility of the excitons. Therefore, EEA is a measure for the exciton mobility \cite{Wolter2017}.
The dynamics can be described by the following rate equation:\cite{Engel2006}
\begin{equation}
    \frac{dN(t)}{dt} = Exc(t)-k_1\cdot N(t) - \frac{1}{2}\gamma\cdot N^2(t) \qquad .
    \label{eq:rate1}
\end{equation}
\textit{N(t)} is the time dependent exciton density, $Exc(t)$ the generation rate due to the excitation pulse, exhibiting a duration of 120~ps (FWHM), $k_1$ the rate for the exponential decay and $\gamma$ the bimolecular annihilation constant. 
The annihilation constant depends on the diffusion constant $D$ and the effective interaction distance (capture radius) $R_{\rm c}$ via\cite{Engel2006}
\begin{equation}
\label{eq:rateEEA}
    \gamma=8\pi D R_{\rm c}
\end{equation}
Here, $R_C$ is calculated as 1.14~nm. It is the hypothetical edge length of a cube with the same volume of an average molecule in the unit cell, i.e. 1.48~nm$^3$.
The diffusion constant is extracted from a series of measurements with increasing excitation flux. 
Photodegradation affects this series significantly. We address it by adapting k$_1$,
see section S5.1 in the ESI for details. The fitted decay is shown in \Fig{\ref{fig:FLIM_timetrace}}.
The model fits the data accurately. We determine the diffusion constant for three different crystals, labelled A, B and C. For crystal A, three areas of the crystal are examined. For the crystals B and C this is done for two and one area, respectively. The exciton diffusion length $L_D = \sqrt{6D\tau}$~~\cite{Mikhnenko2015,Engel2006} ranges from 80 to 180~nm.
The results of the fitting procedure are shown in \Tab{\ref{tab:FLIM_results}}.
From the standard deviation therein we estimate a relative measurement uncertainty of 40\% for $\gamma$ and $D$.
The error of the lifetime is 60~ps. \\
As already discussed, the lifetime of the Frenkel exciton in the crystal can strongly vary at different places inside the crystals. This variation correlates with the amount of photodegradation. Surprisingly, we do not find significant differences in the diffusion constant between these areas. Also, the annihilation constant is not affected by an increasing photodegradation. Thus, the impact of photodegradation is completely covered by adapting the lifetime. The exciton diffusion length on the other hand is by definition affected by the lifetime since it gives the average displacement of an exciton due to diffusion within its lifetime.

\subsection*{Analysis of Exciton Diffusion}
The fluorescence behaviour is determined by radiative and non-radiative decay into the ground state, exciton migration, and EEA. The homogeneous rate equation, Eq. \ref{eq:rate1}, is based on Smoluchowski theory, which assumes that two excitons undergo relative diffusion and annihilate once being within the capture radius $R_{\rm c}$. In this diffusion limited regime the EEA rate, Eq. \ref{eq:rateEEA}, encodes exciton transport ($D$) and their interaction range $(R_{\rm c})$. Further note that Eq. \ref{eq:rateEEA} describes homogeneous diffusion in three dimensions. \\
In what follows we will use KMC simulations to rationalize the results in Table \ref {tab:FLIM_results} in terms of a microscopic picture. To justify the incoherent hopping approach, we note that the largest rate is calculated to be $k_{\rm DA} =1.3~{\rm ps}^{-1}$, it connects monomers 2(4) and 6(8).
In the analysis of the transient absorption spectra given in the next section, a 200 fs decay component will be assigned to vibrational relaxation of the initially excited state. If compared to the F\"orster rate this suggests that indeed the conditions for validity of F\"orster theory (local relaxation faster than transfer) are fulfilled.\cite{may23} \\
For the radiative and non-radiative relaxation to the electronic ground state, an exciton decay rate of 2.5~$\rm ns^{-1}$ is used, calculated from the experimental lifetime of 400~ps, i.e. the lifetime without EEA.
The actual EEA events are treated as an ultrafast process compared to exciton hopping. Therefore, an effective rate constant for EEA of 1~$\rm fs^{-1}$ is assumed, ensuring that annihilation occurs almost instantaneously once two excitons occupy the same lattice site. If compared to the diffusion theory, this implies that $R_{\rm c}$ corresponds to the nearest neighbour distance. \\
Note that the obtained electronic couplings were uniformly scaled by a factor of 1.35, accommodating e.g. uncertainties with respect to the spectral overlap. A single factor preserves the directional trends and relative values of the couplings, while improving agreement with the measured exciton transport behaviour in Fig. \ref{fig:FLIM_timetrace}. 
\begin{figure}[b]
\centering
  \includegraphics[width=\linewidth]{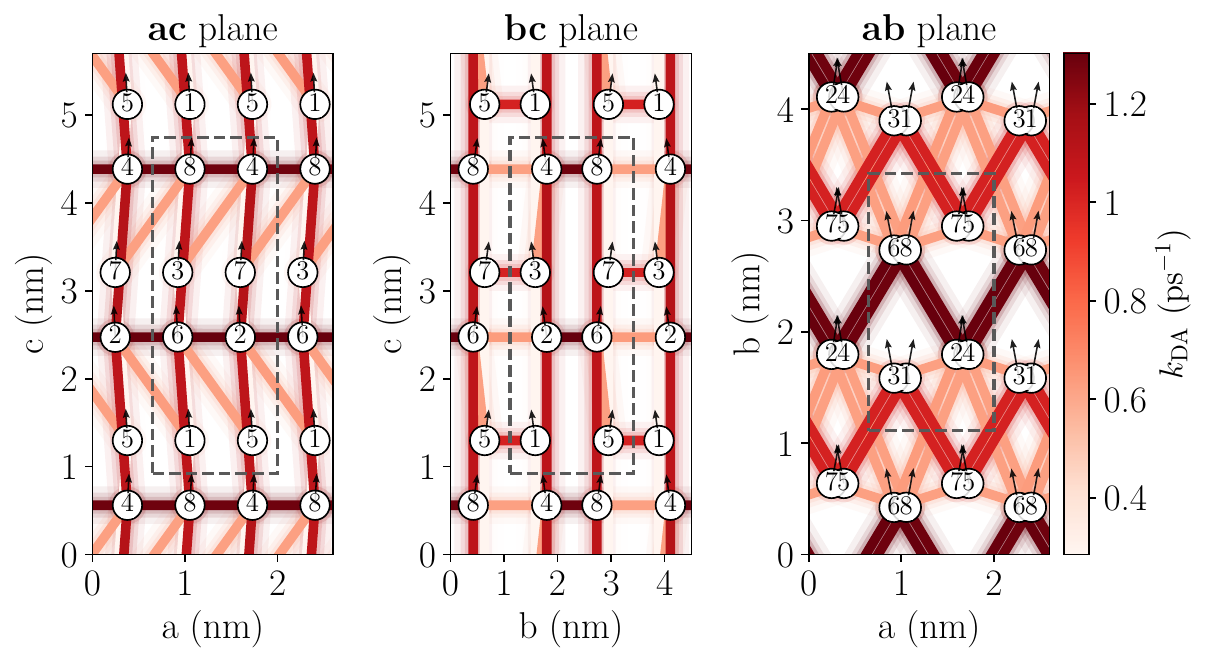}
  \caption{F\"orster rates $k_{\rm DA}$ connecting different monomers for different  crystal planes as indicated. The unit cell is shown (dashed rectangle) together with the nearest neighbors. The patterns of coupling strengths and distances is shown in Fig. S20 and S21, respectively.}
  \label{fig:ratepattern}
\end{figure}
The results of the KMC simulation are shown  in Fig.~S18. Altogether, assuming incoherent hopping of the excitons  fits the experimental annihilation behaviour at both densities  very well.
A mean squared displacement (MSD) analysis is performed by averaging over a number of excitons and individual KMC trajectories in Fig. S18(b).  
These results highlight two consistent trends. First, the acceleration in the decay of the exciton population at higher exciton densities  indicates dominant EEA kinetics, i.e. the exciton count drops rapidly in the case of the higher density. 
Second, the exciton transport is essentially density-independent, showing almost identical linear MSDs which yield the diffusion coefficients $\rm 3.48 \pm 0.35~nm^2/ps$ and $\rm 3.30 \pm 0.38~nm^2/ps$ for higher and lower exciton density, respectively. These values are close to the average experimental result of $\rm 3.80~nm^2/ps$. \\
At this point one should recall that the agreement between the Smoluchowski type rate model for homogeneous three dimensional diffusion and the KMC model is not trivial. First of all, for an orthorhombic crystal one would not necessarily expect isotropic diffusion. To scrutinize this point, we have calculated diffusion constants for motions projected onto the different crystal axes, see Fig. S19 in ESI. They are in a comparable range, i.e. diffusion is indeed to a good approximation isotropic. To rationalize this behaviour we show the F\"orster rates connecting monomers for different spatial projections in Fig. \ref{fig:ratepattern}. Indeed this figure illustrates that the  rates are of similar magnitude for the different  spatial directions. Hence we can conclude that the crystal packing leads to an isotropic diffusion to a good approximation. One should note that the dipole approximation does not hold. Therefore the observation that rates  are comparable for different crystal directions, despite the fact that intermonomer distances are rather different, is a nontrivial result of the special packing of monomers in the unit cell (cf. Figs.~S20 and S21 in ESI). Overall, this justifies the use of the homogeneous diffusion model a posteriori. Since both approaches give a similar diffusion constant and $\gamma$ is obtained by fitting the experimental data, the only free parameter in Eq. \ref{eq:rateEEA} is the effective interaction distance $R_{\rm c}$. The fact that both approaches consistently use the nearest neighbour distance, also justifies the assumption of short-ranged EEA.

\subsection*{Ultrafast Dynamics from Transient Absorption Spectroscopy}
\begin{figure}[t]
\centering
  \includegraphics[width=1\linewidth]{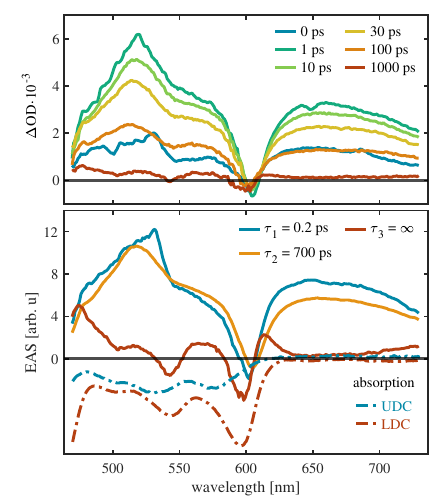}
  \caption{Femtosecond transient absorption spectra of crystal D after photoexcitation at 540~nm. The crystal and the spot size are shown in \Fig{\ref{fig:abs}c}. The exciton density is 5.5$\cdot$10$^5$/µm$^3$. In panel a) transient spectra at different delay times between pump and probe are shown. In panel b) the evolution associated spectra (EAS) along with the LDC and the UDC absorption are depicted. The absorption spectra are plotted to the negative to make a comparison to the bleach easier. Please note that the long-lived signal in the TA data $\Delta$OD($\lambda$) is much smaller than the amplitude of its EAS seems to suggest as at long times the exciton density is very low.}
  \label{fig:TA_data}
\end{figure}

The initial dynamics in the PR crystals after an absorption event is studied by femtosecond transient absorption (TA) spectroscopy since the time resolution of the FLIM measurements is insufficient to monitor the first relaxation steps. To this end a pump-probe microscope is used which allows to study a single crystal, for details see above.
The crystal is excited by laser pulses with a centre wavelength of 540~nm and a duration of 30~fs and the induced TA signal is probed by a white light continuum. 
Pump and probe are polarized parallel to the {\bf b}-axis of the crystal, i.e. they interact with the LDC of the system.
Obtained TA spectra are depicted in \Fig{\ref{fig:TA_data}}a). 
They exhibit a broad excited state absorption (ESA) which covers the entire probed spectral range and which is superimposed by a strong negative feature with a dip at about 605~nm. 
This feature results from a combination of ground state bleach (GSB) and stimulated emission (SE) as a comparison with the stationary LDC absorption shows. 
The TA spectra exhibit some spectral reshaping on the sub-picosecond time scale and an almost complete decay of the signal within several hundred picoseconds. 
Only a very weak absorption change remains, which persists longer than the experimental time window of 1.9~ns. \\
To analyze the data, a rate model is globally fitted to the TA spectra revealing the spectra and lifetimes of the involved species. 
It consists of an ultrafast relaxation step, described by the time constant $\tau_1$, from the optically populated to a relaxed excited state species, which has a lifetime $\tau_2$ of several hundred picoseconds, and a long-lived component with an infinite lifetime. 
Since a rather high exciton density of 5.5$\cdot$10$^5$/µm$^3$ is applied to obtain a good signal-to-noise ratio, EEA needs to be included in the rate model. The annihilation constant $\gamma$ is thereby taken from the FLIM measurements. 
Details of the fit procedure are given in section S5.2 of the ESI. It turns out that the applied fit model describes the TA data very accurately.
The resulting evolution associated spectra (EAS), corresponding to the TA spectra presented in \Fig{\ref{fig:TA_data}}a), are shown in \Fig{\ref{fig:TA_data}}b) and therein compared to the negative stationary LDC and UDC absorption spectra, which should reflect GSB contributions if probed with the corresponding probe polarization. 
The obtained time constants are $\tau_1 = 0.2~$ps and $\tau_2 = 700~$ps. 
Because of EEA, the actual signal decay is faster than $\tau_2$ indicates. 
The EAS of the two decay components look similar and their differences correspond essentially to some spectral reshaping. 
We assign the ultrafast step therefore to vibrational relaxation processes, which are induced by the vibrational excess energy launched by the pump pulses at 540~nm, i.e. well above the absorption edge. 
After this step, the population is in the vibrationally relaxed, lowest electronically excited singlet state. 
The corresponding EAS, i.e. the one associated with $\tau_2$, reflects the dominant TA contributions, i.e. the broad ESA band superimposed by GSB and SE. 
The SE is typical for Frenkel excitons while excimer or charge-transfer excitons would show no significant SE and would exhibit either a characteristic broad (excimer)\cite{Rehhagen2022,Ramirez2020,Brown2014} or a red-shifted (charge-transfer)\cite{Rehhagen2022,Kim2022,Sebastian2021,Wu2015} ESA.
Thus, we identify the species responsible for the $\tau_2$-component as singlet Frenkel excitons. \\
These findings confirm the FLIM analysis and the ab-initio results as the description of the excited state species as singlet Frenkel excitons agrees with the observed fluorescence spectrum and the calculated electronically excited states. 
Furthermore, the applied rate model is, beside of the ultrafast relaxation step, identical with that used for the analysis of the FLIM data and the obtained value for $\tau_2$ agrees with the exciton lifetimes found by FLIM. 
Finally, the TA measurements show that relaxed excitations form within a few hundred femtoseconds and thus much faster than the duration of a hopping step justifying the applied exciton hopping model. \\ 
The EAS of the long-lived component observed in the TA experiments is dominated by GSB and exhibits a weak ESA, which is different from that of the Frenkel excitons while SE is missing.
Thus, it cannot be due to Frenkel excitons. 
We suggest it is related to the population of energetically lower lying trap states associated with crystal defects. 
The latter are probably those that are also responsible for the spatially varying exciton lifetime observed by FLIM, see \Fig{\ref{fig:FLIM}}. 
Their contribution to the TA data is actually very small as can be seen in \Fig{\ref{fig:TA_data}}a) from the TA spectrum at a delay time of 1000 ps. 
Only a small fraction of the excitons is trapped by the defect states while most of them return directly back to the ground state. 
This is accounted for in the rate model by EEA which is the dominant decay channel and deactivates the affected excitons to the ground state.

\section*{Conclusions}
In this work, we investigated the stationary and dynamical properties of excitons in a molecular crystal made up of a large unit cell containing eight PR molecules. 
The dimensions of the unit cell result from the fact that it contains two tetrameric clusters, which are slightly tilted relative to one another and consist of four slipped-stacked PR monomers.
The absorption spectra of the crystals exhibit two perpendicularly polarized components.
In the literature, this observation is typically connected to a translationally nonequivalent pair of monomers in the unit cell, resulting in the splitting of the excited state into two states with non-vanishing transition dipole moment.
In contrast to that, DFT calculations based on the crystallographic data show that in the present case the splitting results from the interaction between the clusters and not from a single monomer pair. Specifically, the observed transitions do not belong to the same Davydov pair, i.e. the observed splitting is not directly connected to a specific element of the coupling matrix.
The calculations predict three transitions with a significant dipole moment. Two of them can be seen in the absorption spectra and are responsible for the two perpendicularly polarized components. The third and strongest one does not show up in the absorption measurements since, due to the orientation of the crystals, its transition dipole is perpendicular to the incident light field. \\
We find highly mobile Frenkel excitons using femtosecond transient absorption microscopy and fluorescence lifetime imaging.
The mobility is much higher than in amorphous nanoparticles consisting of the same molecule.\cite{Rehhagen2022}
In contrast to the nanoparticles, the molecular arrangement in the crystal allows for J-like absorption features.
They lead to an increased spectral overlap, which results in an increased exciton mobility.
Kinetic Monte Carlo simulations based on the calculated coupling matrix and the experimentally determined spectral overlap show a very good agreement with the experimentally observed exciton dynamics for different initial exciton densities, thus covering exciton-exciton annihilation as well. \\
In summary, our work on the excitonic properties of organic single crystals shows that the combination of a crystallographically determined microscopic structure with theoretical calculations and stationary and time-resolved spectroscopic measurements leads to an in-depth understanding of the exciton dynamics and a quantitative validation of the theoretical models. This paves the way for a targeted design of the excitonic properties of organic systems.

\section*{Author contributions}
CR, MF \& SL were responsible for the experimental part and TA \& OK for the theoretical. The manuscript was written by both groups together. \\
CRediT authorship contribution statement: \\
C. Rehhagen: Conceptualization (supporting), Methodology, Software, Validation, Formal analysis, Investigation, Data curation, Writing — original draft (lead), Writing — review \& editing (supporting), Visualization, Supervision (supporting), Project administration (supporting). \\
T. Abdurakhmonov: Investigation, Writing — original draft (supporting), Writing — review \& editing (supporting), Visualization. \\
M. Frank: Validation (supporting), Investigation (supporting), Visualization (supporting). \\
O. Kühn: Conceptualization (lead), Resources (lead), Writing — original draft (equal), Writing — review \& editing (lead), Supervision, Project administration, Funding acquisition. \\
S. Lochbrunner: Conceptualization (lead), Resources, Writing — review \& editing, Supervision, Project administration, Funding acquisition.



\section*{Acknowledgements}
This study was funded by the Deutsche Forschungsgemeinschaft (DFG, German Research
Foundation) via the CRC 1477 “Light–Matter Interactions at Interfaces” (project no. 441234705)
and the IRTG 2676 ”Imaging Quantum Systems” (project no. 437567992). We thank
Alexander Villinger (University of Rostock) for performing the X-ray diffraction measurements.



\balance


\bibliography{refs_exp_theo} 
\bibliographystyle{rsc} 


\section*{Supplementary material}

# S1 Sketches of Experimental Methods

Fig. S1, S2 and S3 show sketches of the experimental set-ups for absorption and emission microscopy and femtosecond pump-probe microscopy. In Fig. S4 an image of an investigated crystal together with the probe spot is shown.

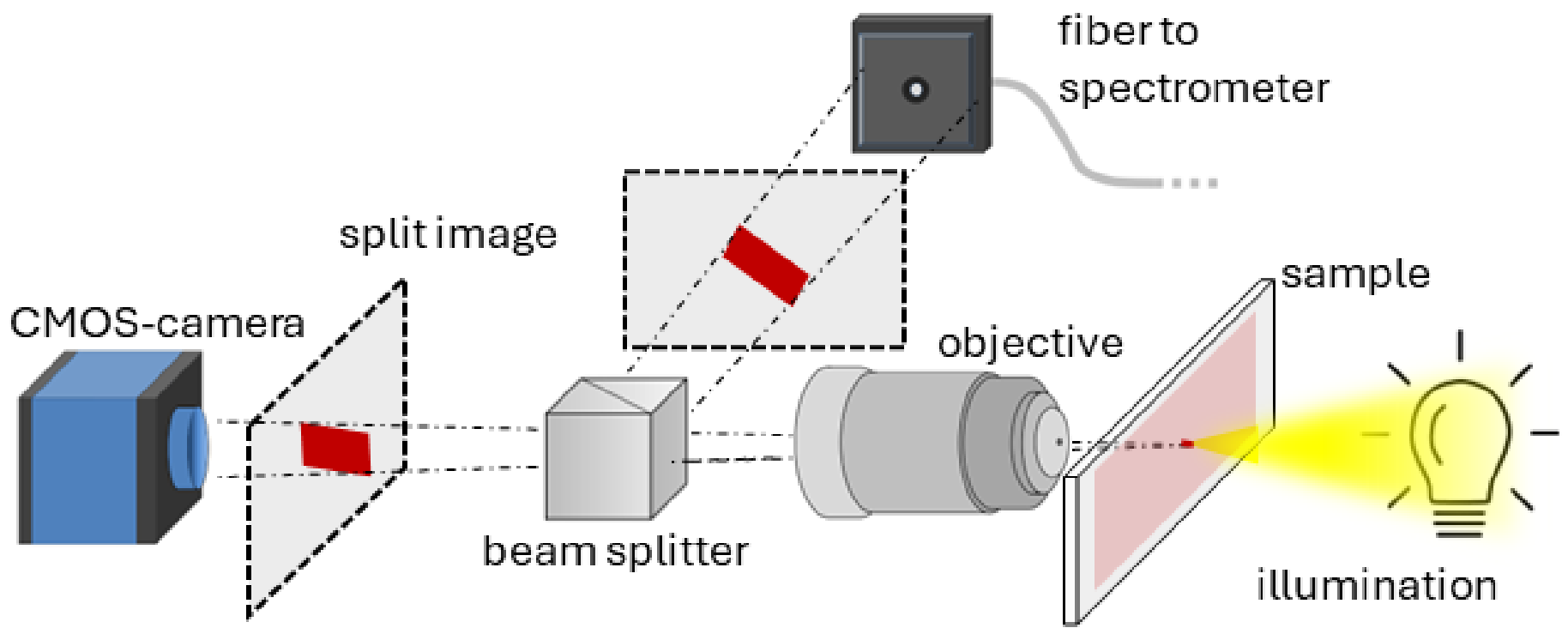


Figure S1: Experimental setup for absorption microscopy.

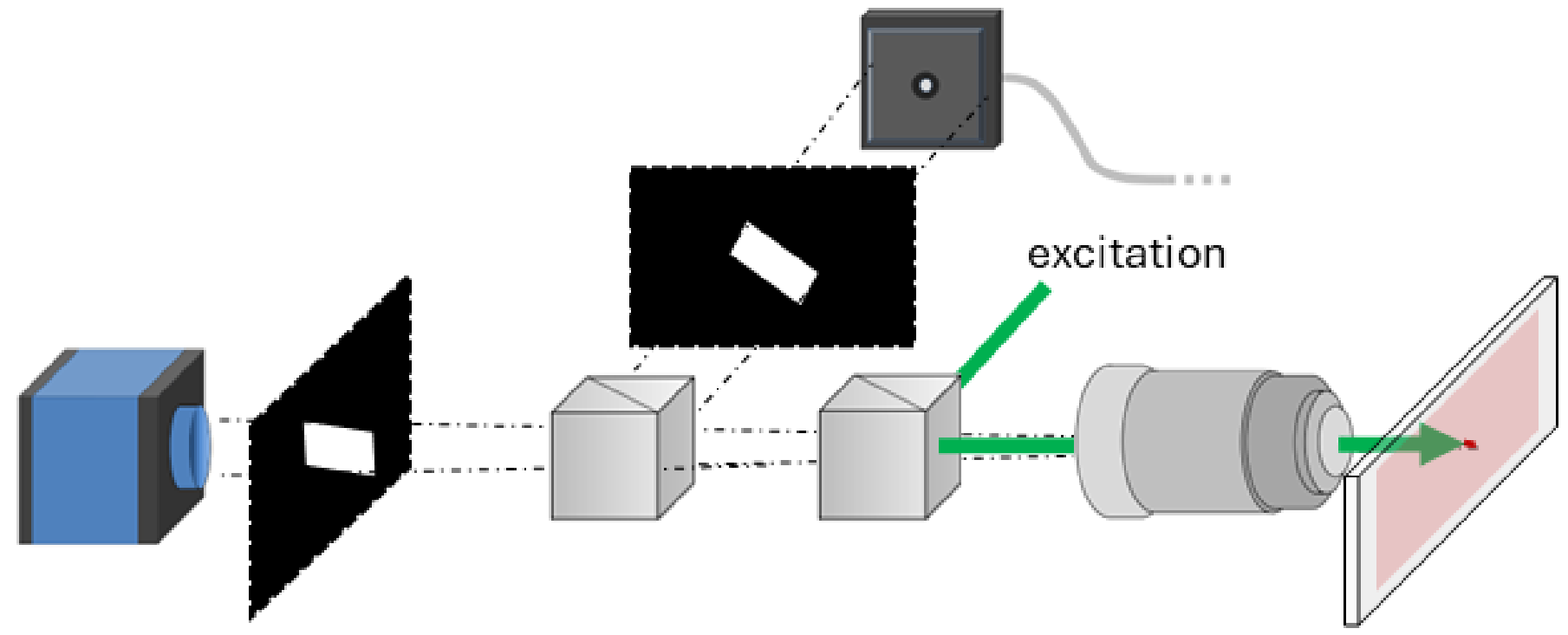


Figure S2: Experimental setup for emission microscopy.

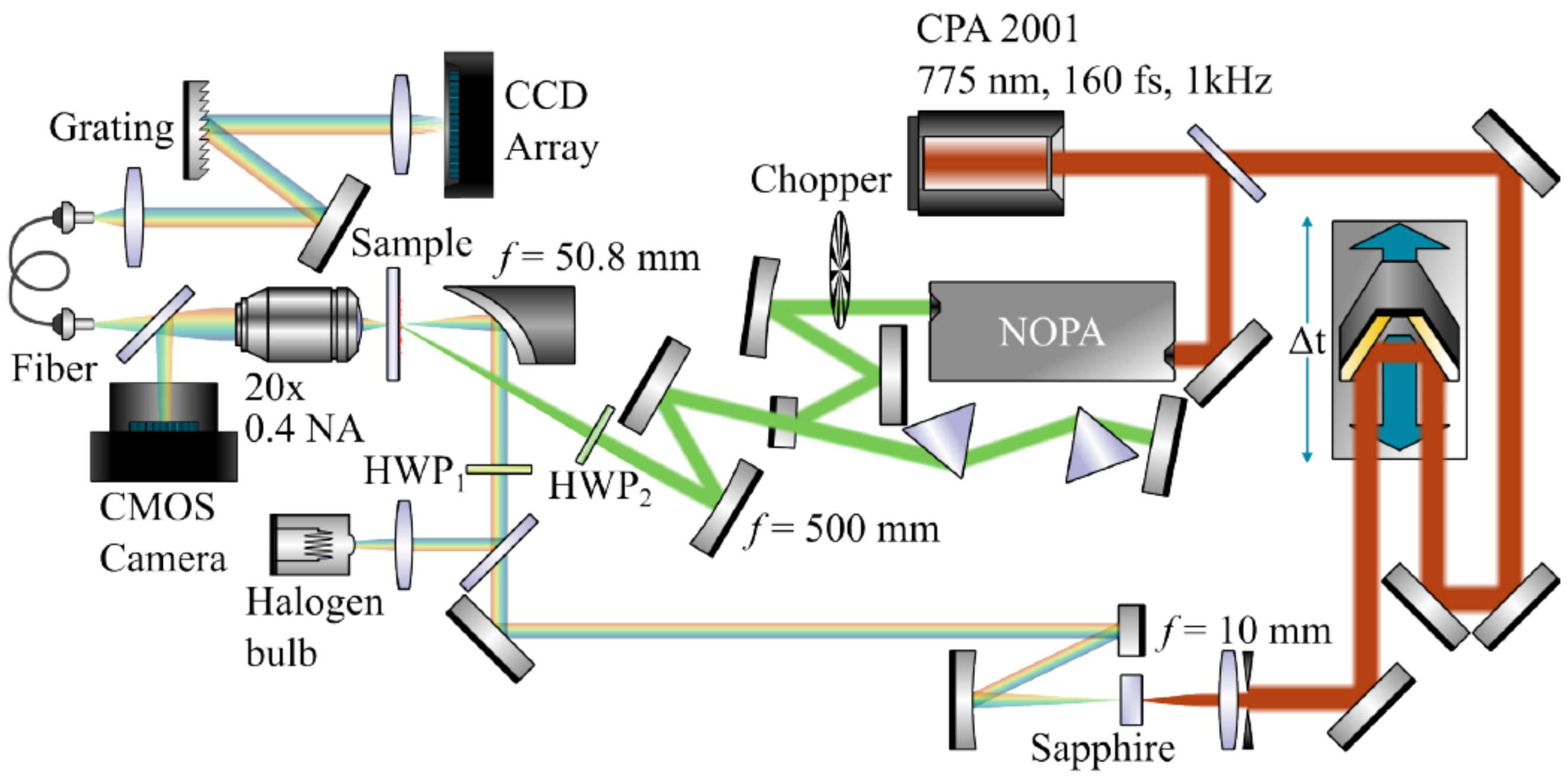


Figure S3: Experimental setup for pump-probe microscopy. *HWP* stands for *half-wave plate*. See description in the main manuscript for further explanation.

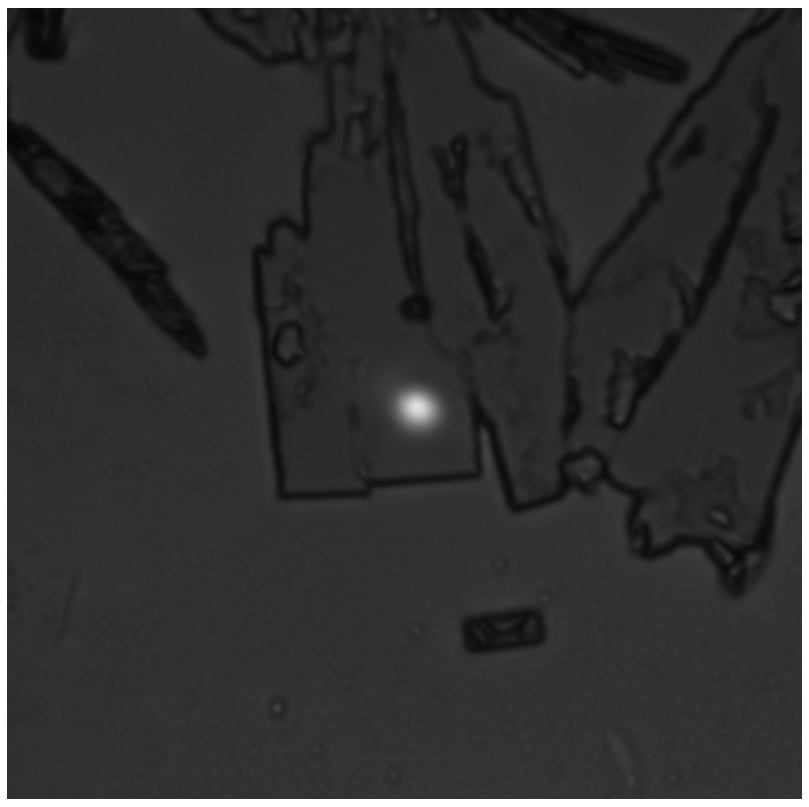

Figure S4: Microscope image recorded with the pump-probe microscope. The bright spot is the probe focus.

## S2 Additional Images of the Crystals and their Unit Cell

In Fig. S5 transmission and confocal microscopy images of all investigated crystals are shown.

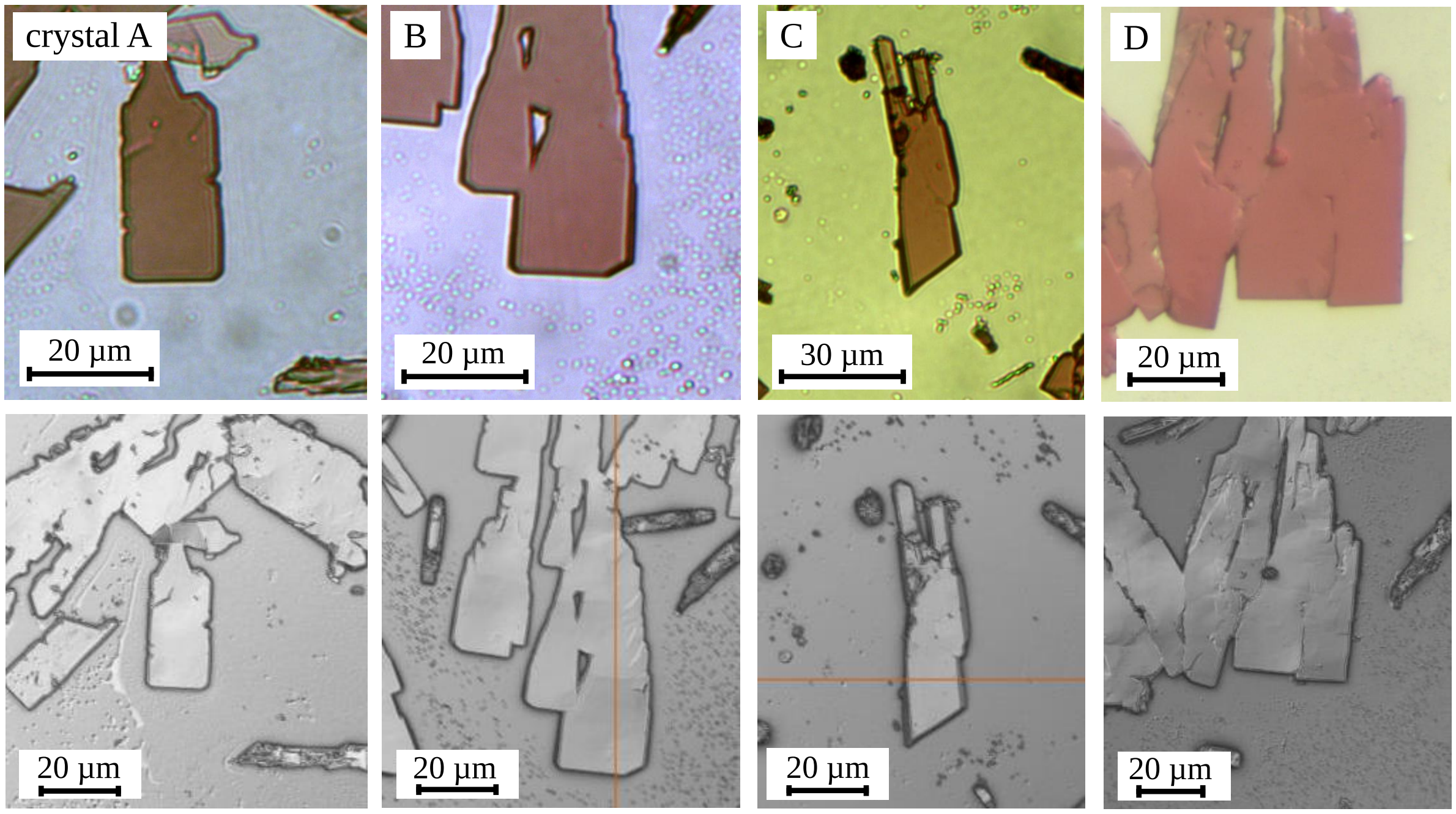


Figure S5: Optical (upper) and confocal microscopy images (lower panel) of the crystals.

In Fig. S6 to Fig. S10 the unit cell is shown from different perspectives and in two versions. In one version all atoms except the hydrogen atoms are depicted and in the other the substituents at the bay and the imide positions are omitted.

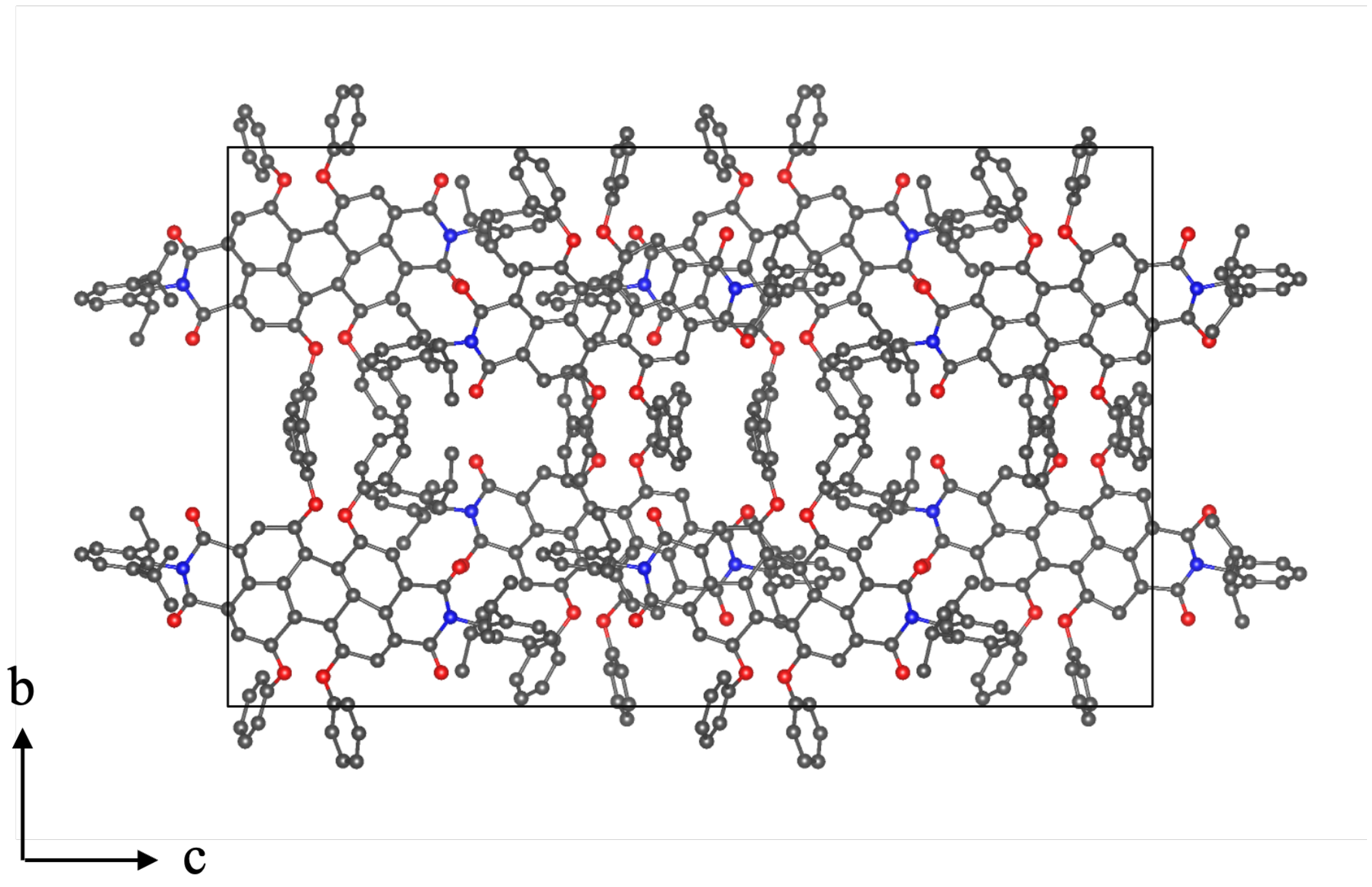


Figure S6: Unit cell along the a-axis.

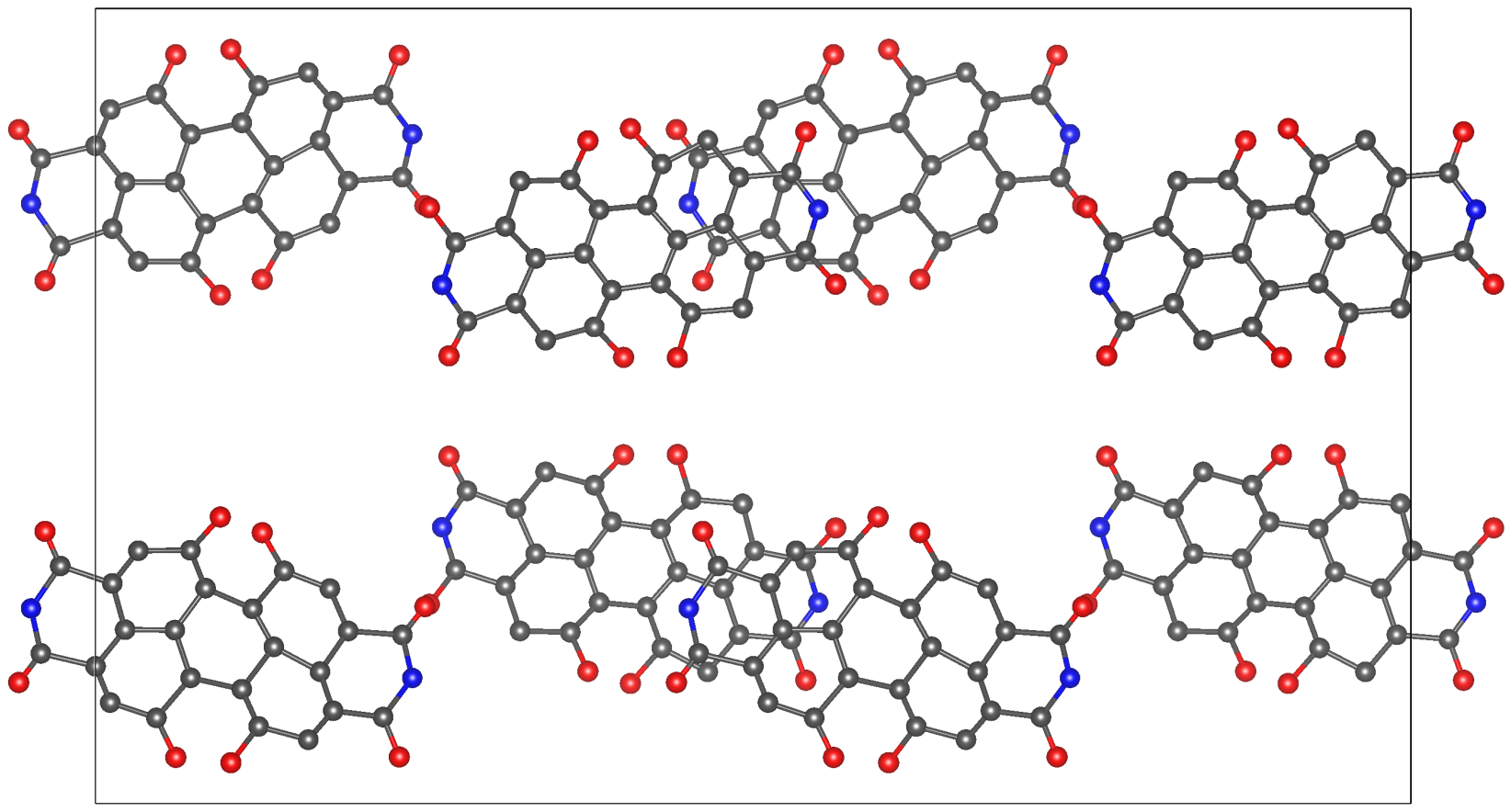

Figure S7: Unit cell along the a-axis. For simplicity, the substituents at the bay and the imide positions are omitted

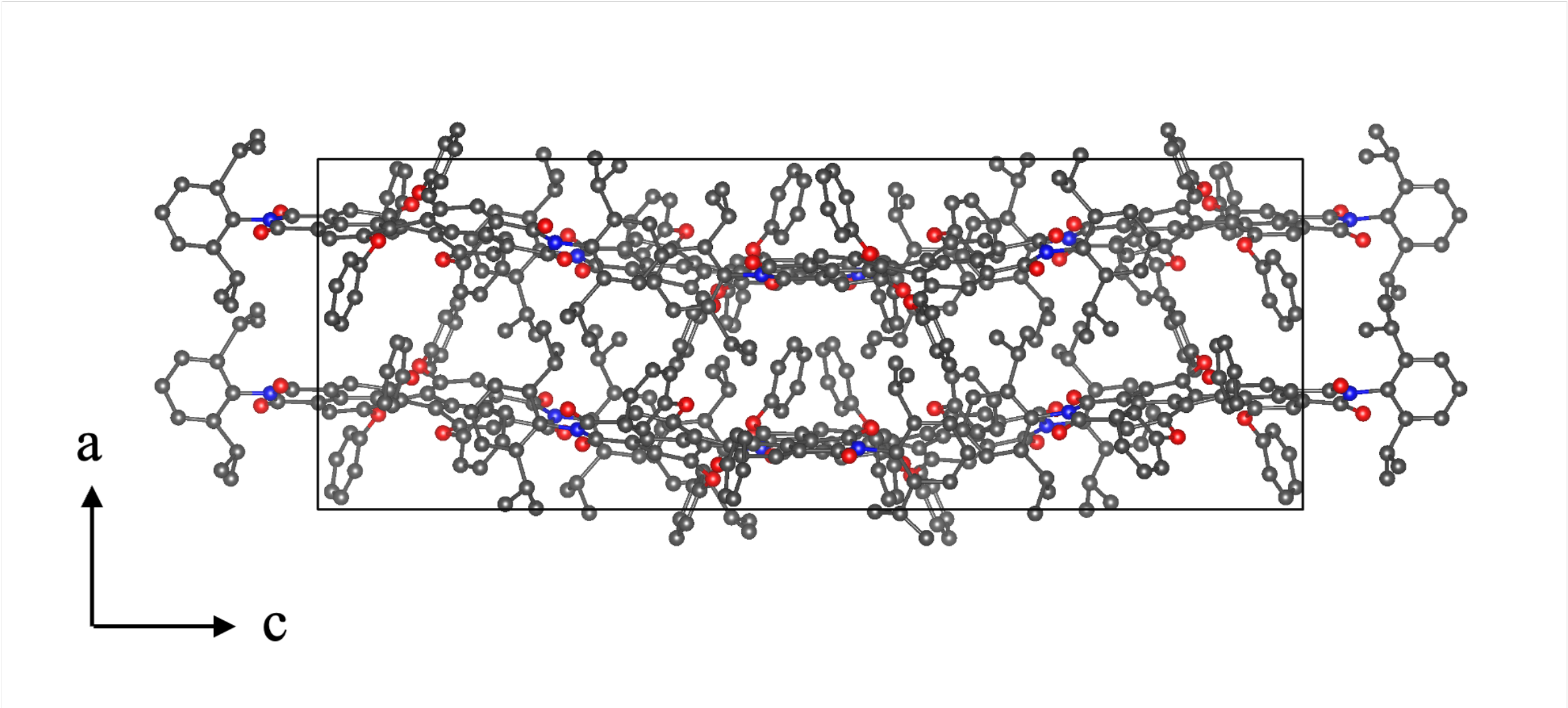


Figure S8: Unit cell along the b-axis.

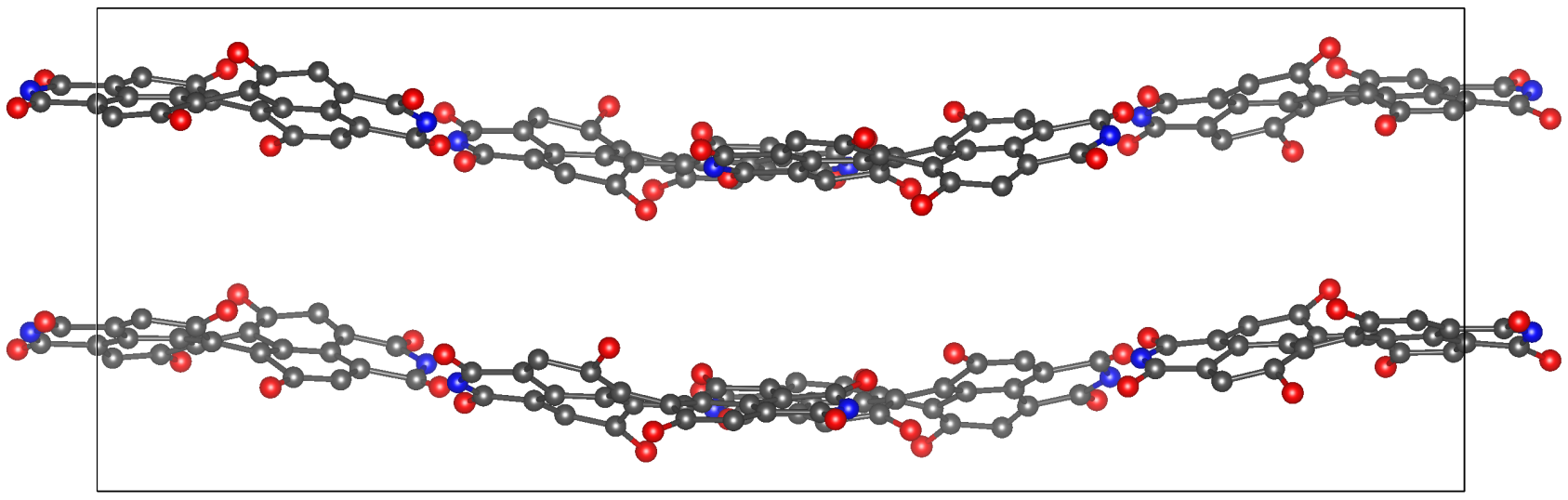

Figure S9: Unit cell along the b-axis. For simplicity, the substituents at the bay and the imide positions are omitted.

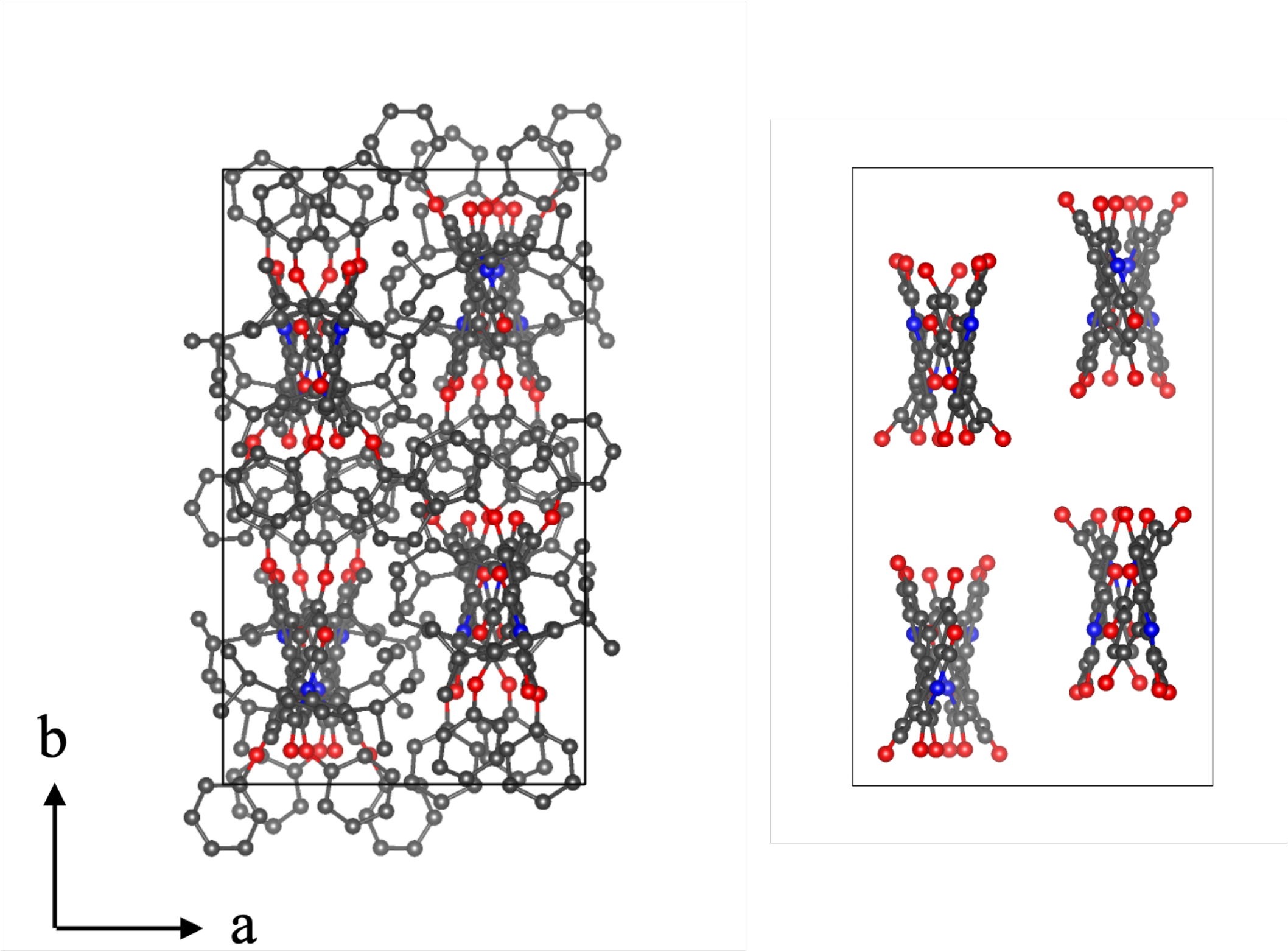


Figure S10: Unit cell along the c-axis. In the right figure, the substituent at the bay and the imide position are removed for simplicity.

# S3 Single Crystal Emission Spectra

The emission spectrum of a single crystal is shown in Fig. S11 in comparison to the monomer fluorescence in solution. The emission is subject to strong self-absorption resulting in a reshapening of the spectrum and dominant contributions in spectral regions without crystal absorption. This can be seen by comparing the emission in transmission and reflection geometry. Since the intensity of the excitation light decays exponentially in the crystal, so does the exciton density. Below an OD of 0.1 this effect is negligible, resulting in a quasi-homogenous distribution of the excitons. For higher OD, the exponential decrease in exciton density with increasing depth has to be taken into account. As the exciton density is highest close to the irradiated crystal surface, the majority of the emission results also from this region. For emission measurements in transmission geometry, the detected emission must therefore, in average, travel a longer distance through the crystal than in case of reflection geometry. Accordingly, the effect of self-absorption is larger in transmission than in reflection. This feature is visible in Fig. S11.

For calculating the Förster exciton transfer rate, the undisturbed emission spectrum is needed to obtain the correct spectral overlap. The undisturbed emission spectrum is reconstructed as follows: First, the emission spectrum obtained in reflection geometry is divided by the absorption in order to estimate the spectral position of the 0-0-band of the emission, see black curve in Fig. S11b). Then, the line shape functions are calculated from the absorption spectra for both the corrected emission and the absorption. Thereafter, a mirror image of the absorption line shape function is generated. It is then shifted to the position of the corrected emission spectrum, see light blue curve in Fig. S11b). According to Strickler-Bergs-symmetry the resulting spectrum serves as an approximation for the undisturbed emission line shape. This procedure is carried out with an x-axis proportional to energy and not with a wavelength axis.

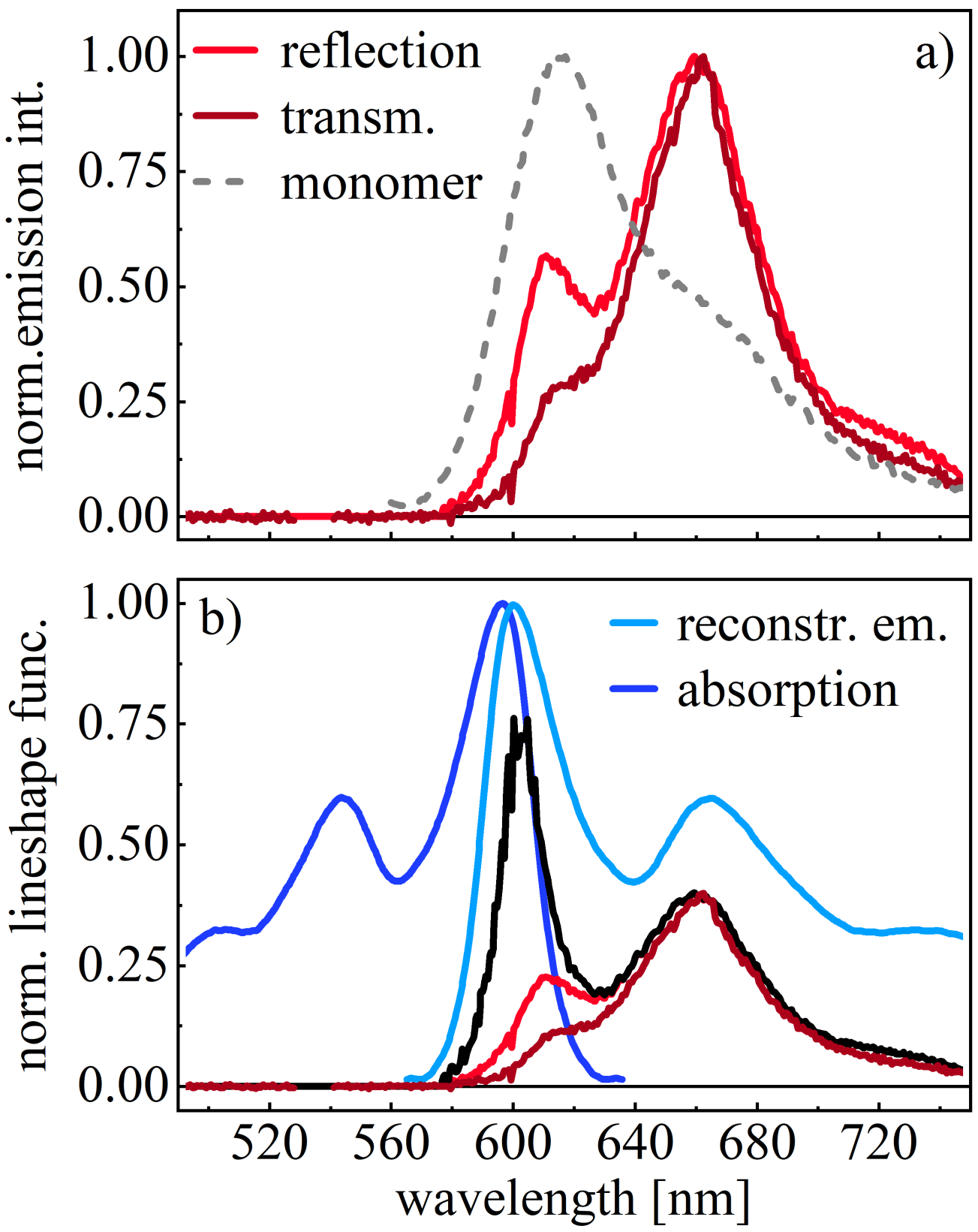


Figure S11: Typical emission spectra of a PR crystal. Panel a): Emission of the crystal measured in transmission and reflection geometry in comparison to the fluorescence spectrum of the monomer in ethanol. Panel b): Reconstruction of the self-absorption-free emission lineshape function of the crystal from its absorption and the measured emission spectrum.

# S4 Photodegradation

The optical and photoinduced properties of the crystals are severely affected by photodegradation. In the solid state it is enhanced compared to thin solution[1] where PR does not show any photodegradation. It manifests itself at high excitation densities, i.e. in certain FLIM and TA measurements, as a reduction of the emission lifetime and the overall emission intensity. The emission peak intensity and the absorption are affected to some extent as well, though not significantly. To monitor the degree of photodegradation, the emission is measured by FLIM before and after a measurement at a very low excitation density. FLIM turns out to be appropriate since it is easy, highly reproducible and sensitive.

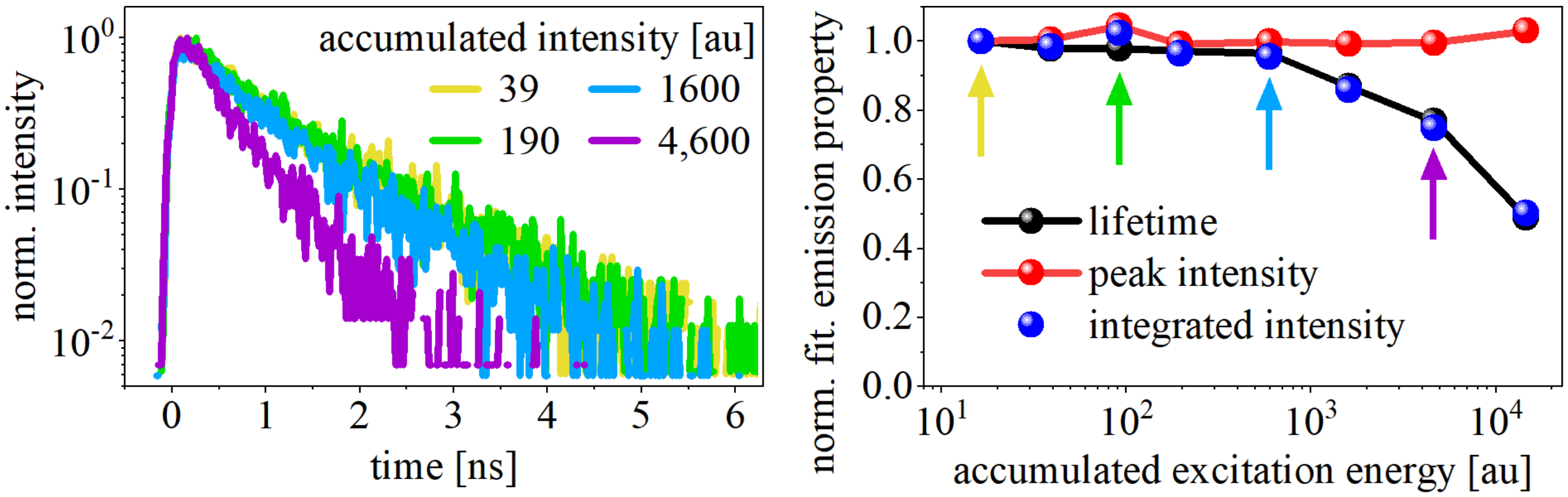


Figure S12: Left: Emission decays of crystal A measured with FLIM at very low excitation density after measurement series with the accumulated excitation energy given in the legend. Please note, that the data shown is taken at an pulse energy of 2.5 au, not at the accumulated energy. Right: Peak and integrated emission intensity and the lifetime extracted from the left panel.

Table 1: Pulse energies for the series of FLIM measurements in Fig. 6 in the main manuscript. The green boxes indicate reference measurements used to track the photodegradation.

| | | | | | | | | | | |
|---|---|---|---|---|---|---|---|---|---|---|
| pulse energy [au] | 1.5 | 2.5 | 10 | 2.5 | 50 | 2.5 | 100 | 2.5 | 400 | 2.5 | 1000 |
| | 2.5 | 3,030 | 2.5 | 10,000 | 2.5 | | | | | | |

For the FLIM measurement series used to extract the exciton mobility, the time-dependence of the emission is measured as such before and after each measurement of the series. Please find this scheme in Tab. 1. It allows to track the development of the photodegradation, which is shown in Fig. S12. In the given example, the lifetime reduces to about 50%. The impact of photodegradation is stronger the longer the original lifetime at the crystal spot under study is.

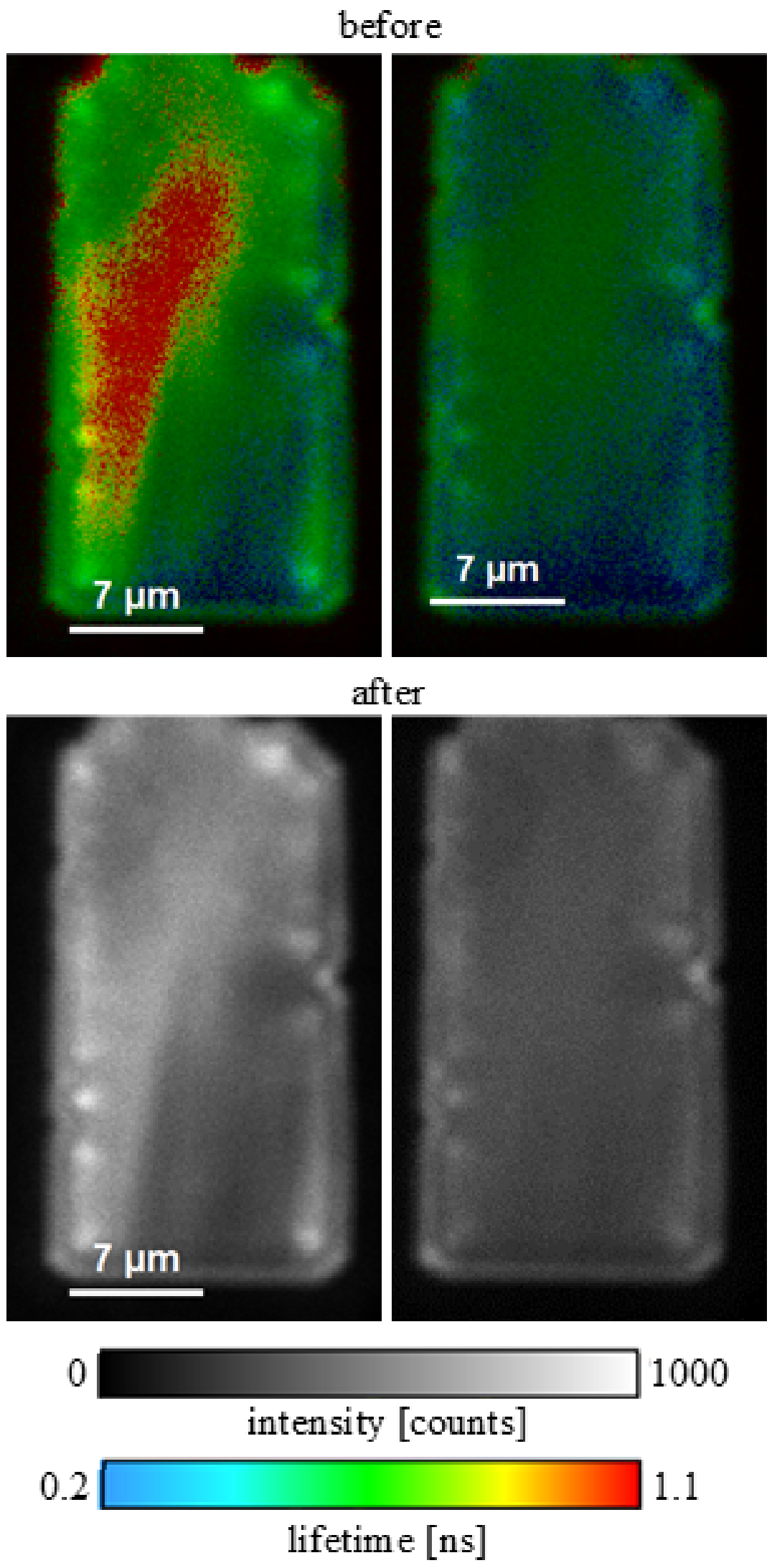


Figure S13: Spatially resolved emission intensity and lifetime distribution in crystal A before (left panels) and after (right panels) the FLIM measurement series. In the upper panel, the lifetime is encoded by the color and the intensity by the brightness. In the lower panel, only the intensity is shown. The scaling is identical.

To determine the exciton mobility from exciton-exciton-annihilation, a series of time-resolved measurements at increasing excitation flux is needed. During this series, the lifetime is continuously reducing due to photodegradation. We account for this problem by tracking the low flux lifetime during this series. The extracted exciton mobility is not notably affected by the reduction lifetime if photodegradation is accurately considered in the analysis. It is worth mentioning, that also the absorption is not affected by photodegradation in the power and measurement time regime applied here. In section S5, it is described how photodegradation is quantitatively taken into account in the analysis of the exciton dynamics.

# S5 Fitting Procedures Applied to the Time-Resolved Measurements

## S5.1 Details on the Fitting Procedure for FLIM Data

The exciton mobility is extracted from the acceleration of the exciton dynamics due to exciton-exciton annihilation. Therefore, a measurement series of the time-resolved emission recorded by FLIM is performed with increasing exciton density, i.e. with increasing pulse energy, see Tab. 1. To analyse the data, the following fit function for the exciton population $N(t)$ is applied to the kinetic emission traces. The fit considers the excitation pulse, an exponential decay and exciton-exciton annihilation:

$$\frac{dN(t)}{dt} = Exc(t) - k_1(E_{\text{acc}}) \cdot N(t) - \frac{1}{2}\gamma \cdot N^2(t) \tag{S1}$$

with the excitation pulse $Exc(t)$ with a duration of 120 ps (FWHM), the decay rate $k_1$ and the annihilation constant $\gamma$.

For the quantitative analysis of the time-resolved emission only a very small area of the crystal is analysed, see Fig. 4 in the manuscript. Therefore, it is valid to assume $k_1$ to be independent on the time although the dynamics of the entire crystal is multi-exponential. Due to photodegradation the lifetime reduces during the series, see section S4. Thus, we define $k_1$ as to be dependent on the accumulated pulse energy $E_{acc}$.

## S5.2 Details on the Fitting Procedure for the TA Data

For the TA-data the following fit function is used:

$$\text{TA}(\lambda, t) = \sum_{i=1}^{N} \text{EAS}_i(\lambda) \cdot n_i(t) \qquad , \tag{S2}$$

with a number $N$ of considered species $i = \{1, 2, \ldots, N\}$, $n_i(t)$ as the time-dependent population of the $i^{th}$ specie and its evolution-associated spectrum $\mathrm{EAS}_i(\lambda)$. The time-dependent evolution of the populations is described with:

$$\frac{\mathrm{d}n_1}{\mathrm{d}t} = -\frac{n_1}{\tau_1} + \frac{n_{exc}}{\sqrt{\pi/\log 16 \cdot \sigma^2}} \cdot \exp\left(-\log 16 \cdot \frac{(t-t_0)^2}{\sigma^2}\right) \tag{S3}$$

$$\frac{\mathrm{d}n_2}{\mathrm{d}t} = +\frac{n_1}{\tau_1} - \frac{n_2}{\tau_2} - \frac{\gamma}{2}n_2^2 \tag{S4}$$

$$\frac{\mathrm{d}n_3}{\mathrm{d}t} = +\frac{n_2}{\tau_2} - \frac{n_3}{\tau_3} \qquad , \tag{S5}$$

with the density $n_{exc}$ of the excitons generated in total, the pulse width $\sigma$, the lifetimes $k_i$ and the annihilation constant $\gamma$.

# S6 Frenkel-Holstein Exciton Model

The absorption spectrum is modelled using the Frenkel-Holstein (FH) Hamiltonian.[2–4] Here local electronic $S_0 \rightarrow S_1$ excitations at site $m$ with energy $\varepsilon_m$ are linearly coupled to a local harmonic vibrational mode. The interaction $J_{mn}$ between local excitations at sites $m$ and $n$ is of resonant Coulomb type. The Hamiltonian is given as follows

$$H = \sum_{m,n}(\varepsilon_m \delta_{mn} + J_{mn})|m\rangle\langle n| + \sum_m \lambda \left(b_m^\dagger + b_m\right)|m\rangle\langle m| + \omega \sum_m \left(b_m^\dagger b_m + \frac{1}{2}\right), \qquad \text{(S6)}$$

Here $b_m^\dagger/b_m$ is the creation/annihilation operator of the vibrational mode at site $m$. Note that the vibrational operators are defined with respect to the undisplaced ground-state oscillator. The displacement of the excited-state potential is introduced through the linear Holstein term, corresponding to a displacement $\lambda/\omega = \sqrt{S}$. We assume that the frequencies, $\omega$, and the vibronic coupling strengths, $\lambda$, are the same for all sites. Due to the relatively large distances between the molecules in the present PR crystals and consequently rather small overlap of the wave functions charge transfer can be neglected.

Transition dipole matrix elements, transition charges, and Coulomb couplings are obtained using linear response time-dependent density functional theory (TDDFT) at the B3LYP/6-31G(d) level. For all TDDFT and Coulomb coupling calculations, the open-source code provided in the PySCF library[5–7] is used. The vertical the electronic excitation energies are considered to be identical for all monomers and taken from the experimental monomer spectrum, $\varepsilon_m = 17316.9$ $\text{cm}^{-1}$.

## S6.1 Benchmark Calculations for the Coupling Elements

There is a plethora of methods to calculate the Coulomb couplings ranging from tight-binding[8] and TDDFT[9] to multi-configurational wave function[10] descriptions. When it comes to larger systems particularly efficient is the mapping of the electronic transition density onto distributed

partial charges, i.e.[8,11]

$$J_{mn} = \sum_{i\in m, j\in n} \frac{q_i^{(m)} q_j^{(n)}}{|\mathbf{R}_i - \mathbf{R}_j|}, \tag{S7}$$

where $q_i^{(m)}$ ($q_j^{(n)}$) are transition charge monopoles at the atoms $i(j)$ at positions $\mathbf{R}_i$ ($\mathbf{R}_j$) of the molecule $m(n)$.

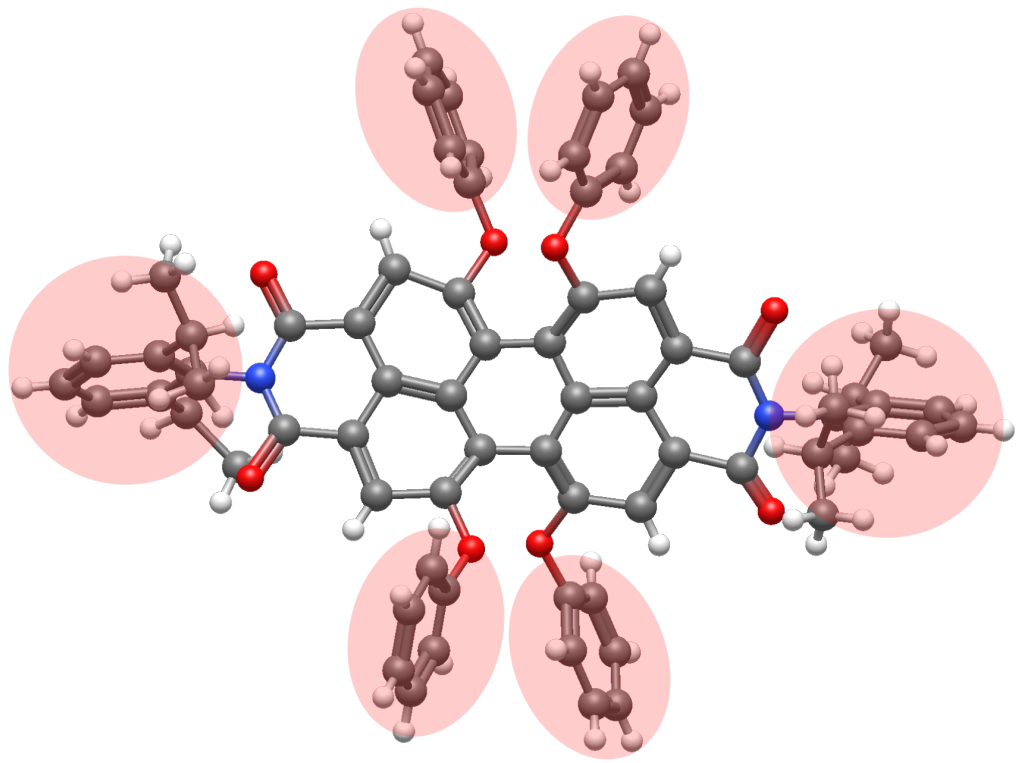

Figure S14: Modified Perylene Red monomer. Highlighted functional groups are removed and terminated with hydrogens to make the calculation more efficient.

To make our benchmark calculations computationally affordable, we only use the perylene core (main nucleus) of the PR monomer to calculate the couplings, because repeating the TDDFT calculation of the complete monomer (138 atoms) for every pair would be very time consuming. All functional groups and heteroatoms highlighted in Fig. S14 are removed since lowest singlet excitations and transition dipole moments are largely localized on the perylene core. In Table 2, it is shown that excitation energies, oscillator strength and transition dipoles of the full monomer and main nucleus are comparable. According to these results, we can even use only the main nucleus to reproduce electronic properties of the complete monomer. The latter is only considered in the calculations for the final absorption spectrum and the exciton dynamics.

Several benchmarks have been performed to find the computationally most efficient yet accurate method. In the first step, we checked by a full TDDFT calculation (Fig. S15a) the contribution of the exchange coupling, which is the short-range part of the excitonic interaction that arises from direct orbital overlap between interacting molecules. Then, three different methods were compared, see Fig. S15(b–d). Since almost all molecules are well-separated (r $\geq$ 10 Å),

Table 2: Comparison of TDDFT results of excitation energies, oscillator strength and transition dipole moments of the complete monomer and the main nucleus.

| Complete monomer | | | | | |
|---|---|---|---|---|---|
| State | Exc.($cm^{-1}$) | Osc. strength | Dip. M. [a.u.] | | |
| S1 | 23138.5 | 0.660 | 3.065 | 0.005 | -0.079 |
| S2 | 26178.9 | 0.005 | 0.103 | 0.185 | -0.130 |
| S3 | 26638.7 | 0.166 | 0.488 | 1.333 | 0.188 |
| S4 | 27211.3 | 0.055 | 0.432 | -0.531 | 0.439 |
| S5 | 27396.8 | 0.004 | 0.140 | 0.146 | 0.112 |
| Reduced Monomer | | | | | |
| S1 | 23582.0 | 0.542 | 2.751 | -0.060 | -0.010 |
| S2 | 27114.5 | 0.085 | 0.463 | 0.894 | 0.136 |
| S3 | 27461.3 | 0.016 | -0.047 | 0.357 | -0.256 |
| S4 | 27864.6 | 0.011 | 0.204 | -0.189 | 0.236 |
| S5 | 28735.6 | 0.001 | 0.096 | -0.005 | 0.039 |

the contribution of exchange coupling can be neglected and we only retain the Coulomb coupling term (Fig. S15a).

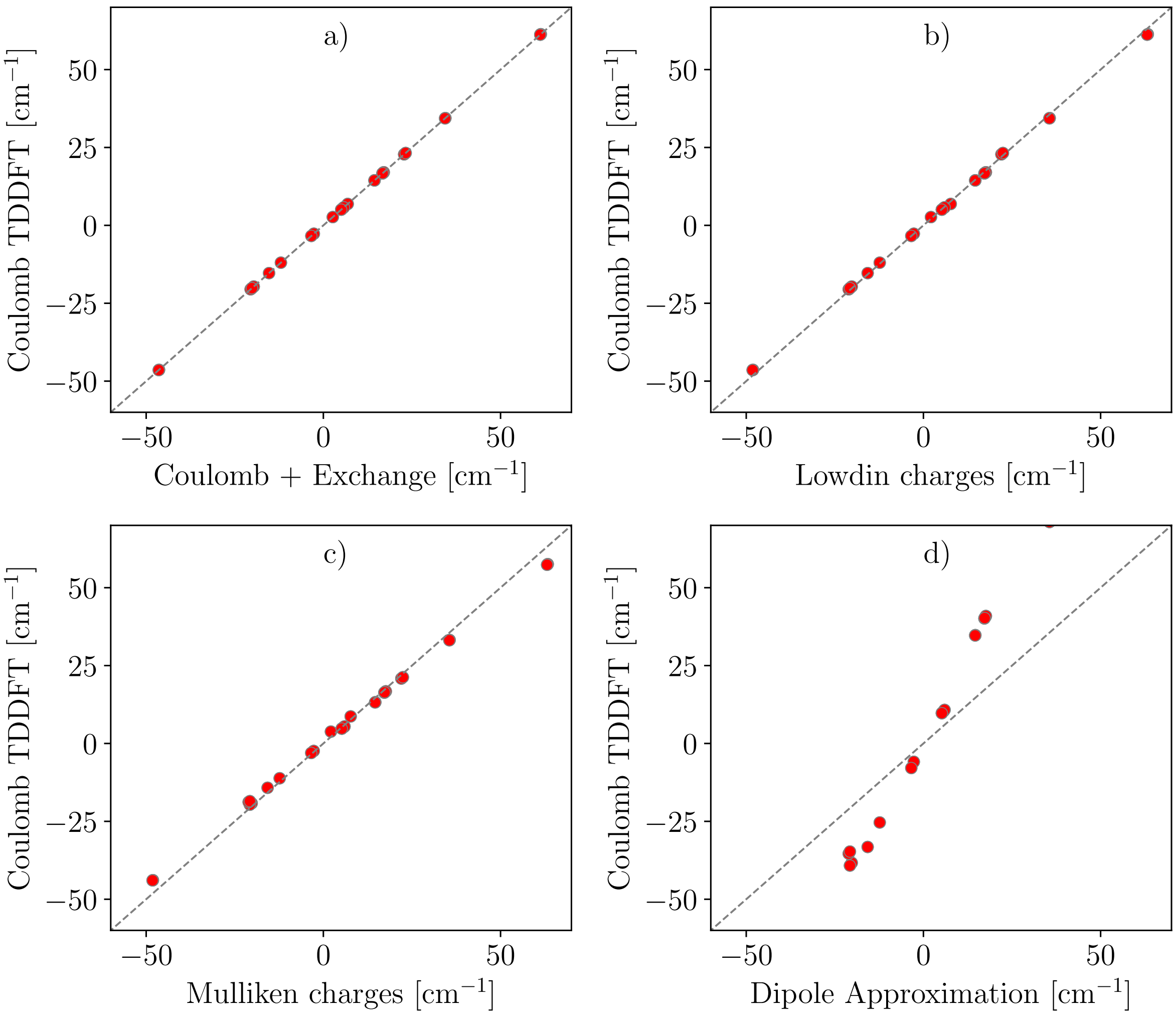


Figure S15: Comparison of coupling calculated by different methods: (a) contribution of exchange coupling, (b) Löwdin charges, (c) Mulliken charges, (d) Dipole-Dipole approximation. There are 28 unique pairs formed by the 8 molecules of the unit cell.

While the often used dipole-dipole approximation severely overestimates the couplings, we found that Löwdin analysis based transition charges performed best (2 % error) and are used in what follows.

## S6.2 Periodic Frenkel-Holstein Model

The periodicity of the crystal is accounted for using a **k**-space (Bloch) representation. The Coulomb couplings are then expressed as

$$J_{mn}(\mathbf{k}) = \sum_{\mathbf{R}}' J_{mn}(\mathbf{R}) e^{i\mathbf{k}\cdot\mathbf{R}} \tag{S8}$$

with $\mathbf{R} = n_a\mathbf{a} + n_b\mathbf{b} + n_c\mathbf{c}$ being the lattice vectors. This expression contains intracell ($\mathbf{R} = 0$) as well as intercell couplings. In the present model, intercell couplings within a cut-off distance of 10 nm are taken into account. The prime indicates that only the self-interaction term $m = n$, $\mathbf{R} = 0$ is excluded. Thus, the sum includes couplings between translationally equivalent molecules in neighbouring cells ($m = n, \mathbf{R} \neq 0$), as well as couplings between nonequivalent molecules ($m \neq n$). Since we are interested in the optical absorption spectrum, only the Γ-point (**k**=0) needs to be considered. Due to the locality of the vibrational modes and the weak to medium strong Coulomb coupling, the vibronic interaction terms in Eq. S6 remain unchanged in the reciprocal space.[3]

Table 3: Excitonic couplings ($J_{mn}$) calculated using transition Löwdin charges. All values are given in cm$^{-1}$. The site index is consistent with the illustration given in Fig. S16.

| $J_{ij}$ | 1 | 2 | 3 | 4 | 5 | 6 | 7 | 8 |
|---|---|---|---|---|---|---|---|---|
| 1 | 0.0 | -21.3 | -29.6 | -7.4 | 22.4 | -25.9 | -4.8 | -6.6 |
| 2 | -21.3 | 0.0 | -5.6 | -58.6 | -26.0 | 63.2 | 27.1 | -16.6 |
| 3 | -29.6 | -5.6 | 0.0 | -43.7 | -4.0 | 26.6 | 22.4 | -26.1 |
| 4 | -7.4 | -58.6 | -43.7 | 0.0 | -6.6 | -12.8 | -26.0 | 62.7 |
| 5 | 22.4 | -26.0 | -4.0 | -6.6 | 0.0 | -43.8 | -29.6 | -7.5 |
| 6 | -25.9 | 63.2 | 26.6 | -12.8 | -43.8 | 0.0 | -8.4 | -57.9 |
| 7 | -4.8 | 27.1 | 22.4 | -26.0 | -29.6 | -8.4 | 0.0 | -21.0 |
| 8 | -6.6 | -16.6 | -26.1 | 62.7 | -7.5 | -57.9 | -21.0 | 0.0 |

The vibrational frequency and the vibronic coupling strength are estimated according to the experimental data as $\omega = 1420$ cm$^{-1}$ and $\lambda = \sqrt{S}\omega = 0.14\ eV$ with $S$ being the Huang-Rhys factor. The latter values are typical for vinyl stretching modes of the perylene core.

The FH Hamiltonian is represented in one-particle approximation,[12] in which only the excited

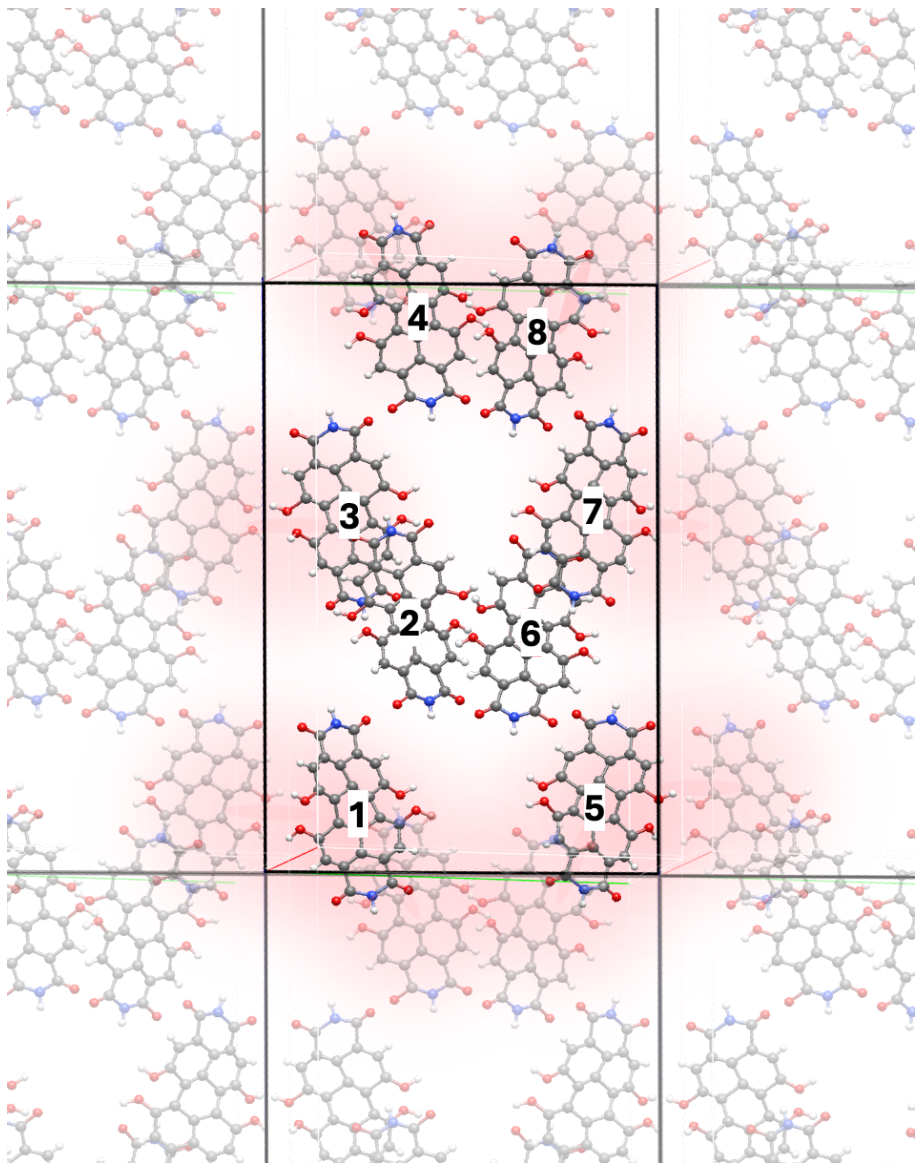


Figure S16: Visualization of PR crystal structure showing the primitive cell and its closest periodic images. Red shaded regions illustrate the strong Coulomb coupling between the molecules inside the unit cell and their closest neighbours in adjacent cells.

vibrational states for the electronically excited molecule are considered. All other molecules are in the electronic and vibrational ground state. The electronic basis contains only one excitation per unit cell while all other chromophores remain in the ground state. The vibronic basis set can be written as $\{|m,\nu\rangle\}_{\nu=0}^{\nu_{\max}}$, where $|m,\nu\rangle$ denotes a local electronic excitation on molecule $m$, together with $\nu$ the undisplaced intramolecular oscillator quanta, and $\nu_{\max}$ is the maximal number of vibrational quanta possible for the electronic excitation at molecule $m$. As described below, we find convergence of the absorption spectrum when using $\nu_{\max} = 5$. Hence, the matrix to be diagonalized is of dimension $N_{\mathrm{cell}}(\nu_{max}+1) \times N_{\mathrm{cell}}(\nu_{max}+1)$ where $N_{\mathrm{cell}} = 8$ is the number of molecules in the unit cell.

We obtain vibronically mixed eigenstates as $|\psi_k\rangle = \sum_{m=1}^{N_{\mathrm{cell}}} \sum_{\nu=0}^{\nu_{\max}} c_{m\nu}^{(k)} |m,\nu\rangle$. Within the Condon approximation, the transition dipole couples the ground state to the $\nu = 0$ component of the undisplaced oscillator basis. The oscillator strength of the $k$-th state is $f_k^{\alpha} \propto \left|\sum_{m=1}^{N_{\mathrm{cell}}} \mu_{m,\alpha} c_{m0}^{(k)}\right|^2$ for a given polarisation direction $\alpha$ ($\alpha$ is $\perp b$ or $\parallel b$). The Franck-Condon progression is contained in the coefficients $c_{m0}^{(k)}$, because the Holstein coupling mixes the vibrational basis states during diagonalisation. The resulting spectrum $\{E_k, f_k^{\alpha}\}$ is convoluted with a Gaussian width of $\sigma =$

55 meV to account for homogeneous broadening.

# S7 Symmetry-based Analysis of the Absorption Spectrum

In the following we will analyse the absorption spectrum given in Fig. S17. Special emphasis is put on the origin of the apparent Davydov splitting of about 520 $cm^{-1}$. Central to the analysis is the observation that the eight monomers in the unit cell can be grouped into two clusters which are (nearly) exchange symmetric.

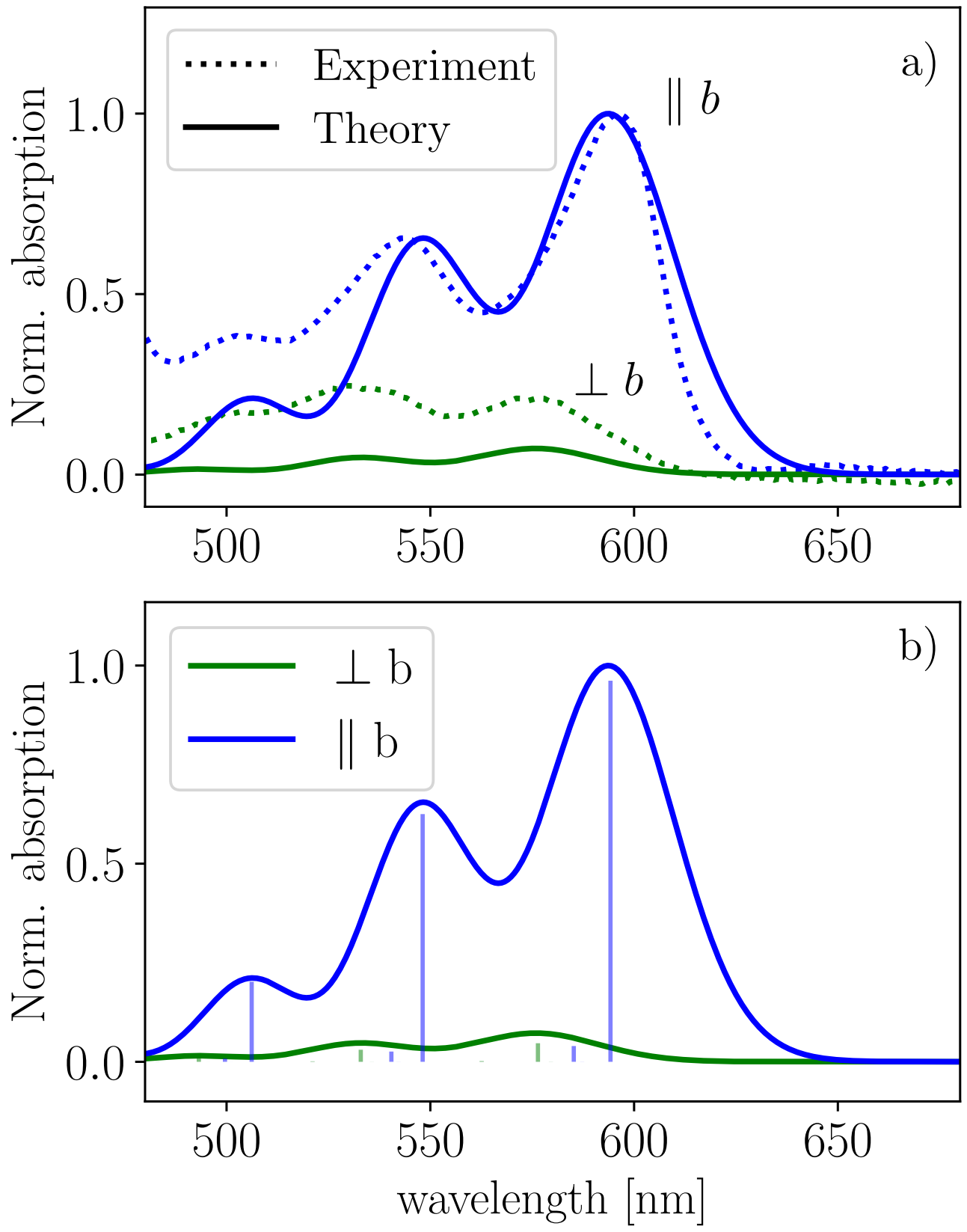


Figure S17: Absorption spectrum for parallel (blue) and perpendicular (green) projections to the **b**-axis of the PR crystal: (a) Comparison between simulated and experimental spectrum. (b) Decomposition of the spectra into stick spectra giving the respective oscillator strengths. The apparent Davydov splitting for the simulated spectrum is $\approx 520$ $cm^{-1}$

.

## S7.1 Analysis of the Hamiltonian Matrix

The Hamiltonian matrix respecting lattice symmetry at $\mathbf{k} = 0$ as calculated using TDDFT derived Löwdin transition charges reads as follows

$$
\mathbf{H} = \begin{pmatrix}
17370.6 & -99.9 & -83.2 & -7.8 & 148.9 & -49.3 & -64.6 & 58.1 \\
-99.9 & 17370.6 & -16.6 & -170.8 & -50.0 & 201.0 & 52.3 & -106.1 \\
-83.2 & -16.6 & 17370.6 & -103.4 & -63.9 & 52.1 & 148.9 & -49.2 \\
-7.8 & -170.8 & -103.4 & 17370.6 & 58.0 & -102.6 & -49.3 & 199.8 \\
148.9 & -50.0 & -63.9 & 58.0 & 17370.6 & -103.6 & -83.2 & -8.2 \\
-49.3 & 201.0 & 52.1 & -102.6 & -103.6 & 17370.6 & -17.7 & -169.7 \\
-64.6 & 52.3 & 148.9 & -49.3 & -83.2 & -17.7 & 17370.6 & -99.1 \\
58.1 & -106.1 & -49.2 & 199.8 & -8.2 & -169.7 & -99.1 & 17370.6
\end{pmatrix} \tag{S9}
$$

Overall it reflects the geometry of the unit cell, i.e. there are two 4×4 blocks (corresponding to geometric clusters or pools) with rather similar entries. This suggests writing

$$
\mathbf{H} = \begin{pmatrix} \mathbf{H}_{11} & \mathbf{H}_{12} \\ \mathbf{H}_{21} & \mathbf{H}_{22} \end{pmatrix} \tag{S10}
$$

In order to quantify symmetry we define the swap operator $\mathbf{I}$ as the 4×4 unit matrix

$$
\mathbf{S} = \begin{pmatrix} \mathbf{0} & \mathbf{I} \\ \mathbf{I} & \mathbf{0} \end{pmatrix} \tag{S11}
$$

Applying the swap operator to the Hamiltonian and comparing with the original one obtains ($||\cdot||$ denotes the 2-norm)

$$
\frac{||\mathbf{S}^{\mathrm{T}}\mathbf{H}\mathbf{S} - \mathbf{H}||}{||\mathbf{H}||} = 2.8 \times 10^{-4} \tag{S12}
$$

We conclude that the 4×4 sectors of the Hamiltonian matrix behave like $\mathbf{H}_{11} \approx \mathbf{H}_{22}$ and $\mathbf{H}_{21}^{\mathrm{T}} \approx \mathbf{H}_{12}$. In other words there is an approximate exchange symmetry with respect to the two clusters.

Next we lock at the block-diagonal approximation, i.e. finding the eigenstates according to

$$\begin{pmatrix} \mathbf{H}_{11} & \mathbf{0} \\ \mathbf{0} & \mathbf{H}_{22} \end{pmatrix} \begin{pmatrix} \mathbf{V}_1 \\ \mathbf{V}_2 \end{pmatrix} = \begin{pmatrix} \mathbf{E}_1 & \mathbf{0} \\ \mathbf{0} & \mathbf{E}_2 \end{pmatrix} \begin{pmatrix} \mathbf{V}_1 \\ \mathbf{V}_2 \end{pmatrix} \tag{S13}$$

with $\mathbf{E}_n$ and $\mathbf{V}_n$ being the diagonal matrix of eigenvalues and the eigenvector for sector $n$. The set of eigenvectors forms a matrix $\mathbf{Q} = \mathrm{diag}(\mathbf{V}_1, \mathbf{V}_2)$ and the full Hamiltonian can be transformed into the basis of the block-diagonal approximation yielding

$$\mathbf{Q}^{\mathrm{T}}\mathbf{H}\mathbf{Q} = \tilde{\mathbf{H}} = \begin{pmatrix} 17121.5 & -0.0 & 0.0 & 0.0 & 97.7 & 3.7 & -0.5 & 1.3 \\ -0.0 & 17365.1 & -0.0 & 0.0 & 2.6 & 83.2 & 4.3 & 4.6 \\ 0.0 & -0.0 & 17398.6 & 0.0 & -2.2 & 5.5 & 144.6 & 3.9 \\ 0.0 & 0.0 & 0.0 & 17597.2 & -0.8 & -5.4 & 5.9 & 373.0 \\ 97.7 & 2.6 & -2.2 & -0.8 & 17121.9 & 0.0 & -0.0 & -0.0 \\ 3.7 & 83.2 & 5.5 & -5.4 & 0.0 & 17365.7 & 0.0 & -0.0 \\ -0.5 & 4.3 & 144.6 & 5.9 & -0.0 & 0.0 & 17399.2 & 0.0 \\ 1.3 & 4.6 & 3.9 & 373.0 & -0.0 & 0.0 & -0.0 & 17595.6 \end{pmatrix} \tag{S14}$$

Inspection of this matrix shows that strong coupling between the cluster eigenstates occurs only for those distinct pairs, which have approximately the same energy, i.e. are near-resonant. Thus as a consequence of the near-exchange symmetry the eigenstates organize in four $\pm$ pairs (each eigenstate of a cluster couples strongly to it's 'counterpart' in the other cluster). This suggests to transform to a cluster-(exchange-)parity basis like $|\pm\rangle \propto (|\mathrm{cluster}_1\rangle \pm |\mathrm{cluster}_2\rangle)$. Formally this is achieved by the transformation which diagonalizes the swap operator (with eigenvalues $\pm 1$ for anti/symmetric states), i.e.

$$\mathbf{Q}_{\mathrm{p}} = \frac{1}{\sqrt{2}} \begin{pmatrix} \mathbf{I} & \mathbf{I} \\ \mathbf{I} & -\mathbf{I} \end{pmatrix} \quad \rightarrow \quad \mathbf{Q}_{\mathrm{p}}^{\mathrm{T}}\mathbf{S}\mathbf{Q}_{\mathrm{p}} = \begin{pmatrix} \mathbf{I} & \mathbf{0} \\ \mathbf{0} & -\mathbf{I} \end{pmatrix} \tag{S15}$$

In the cluster-parity basis the Hamiltonian becomes nearly block-diagonal

$$\mathbf{Q}_{\mathrm{p}}^{\mathrm{T}}\mathbf{H}\mathbf{Q}_{\mathrm{p}} = \mathbf{H}_{\mathrm{p}} = \begin{pmatrix} 17519.5 & -151.4 & -147.4 & 50.0 & 0.0 & 1.5 & 0.4 & 0.1 \\ -151.4 & 17571.6 & 35.0 & -274.6 & 2.2 & -0.0 & 0.4 & 1.2 \\ -147.4 & 35.0 & 17519.5 & -150.5 & -0.3 & 0.6 & 0.0 & -2.2 \\ 50.0 & -274.6 & -150.5 & 17570.4 & 0.2 & -2.3 & -2.1 & -0.0 \\ -0.0 & 2.2 & -0.4 & 0.3 & 17221.7 & -52.1 & -19.0 & -66.0 \\ 1.5 & -0.0 & 0.6 & -2.3 & -52.1 & 17169.6 & -69.3 & -65.9 \\ 0.3 & 0.5 & -0.0 & -2.1 & -19.0 & -69.3 & 17221.7 & -52.0 \\ 0.1 & 1.2 & -2.2 & 0.0 & -66.0 & -65.9 & -52.0 & 17170.8 \end{pmatrix}$$

$$= \begin{pmatrix} \mathbf{H}_{+} & \mathbf{H}_{\pm} \\ \mathbf{H}_{\pm} & \mathbf{H}_{-} \end{pmatrix} \tag{S16}$$

In fact there is little parity mixing

$$\frac{||\mathbf{H}_{\pm}||}{||\mathbf{H}||} = 1.0 \times 10^{-4} \tag{S17}$$

Further analysis is provided by the parity content of the eigenstates of the total $\mathbf{H}$, Eq. (S9), i.e. $\Psi_k$ and $E_k$, defined as

$$p_k = \langle \psi_k | \mathbf{S} | \psi_k \rangle \tag{S18}$$

It turns out that most $p_k$ are close to 1, except two which come as a pair $p_k = \pm 0.82$ ($k = 2, 3$); these are mixed parity states (see Tab. 4). At first glance this looks surprising given the small parity mixing, but the effect could be amplified by close resonance. Indeed checking the eigenvalues of separate $\pm$ sectors of $\mathbf{H}_{\mathrm{p}}$ one finds that there is one pair for which $E_{+,1} - E_{-,2} = 3.3$ cm$^{-1}$.

To summarize this analysis, most states can be classified according to parity, except one pair with close to resonant energies. Notice that this resonance effect would be sensible to small imperfections of the crystal.

## S7.2 Analysis of the Spectrum

Table 4: Analysis of eigenstates, $\Psi_k$ (energies $E_k$ in cm$^{-1}$), and dipole transitions of the full Hamiltonian, Eq. (S9). The dipole transitions ($\mu$ given in atomic units) are classified according to their even ('e') and odd ('o') character, which is further specified according to the different spatial components as indicated.

| $k$ | $E_k$ | $p_k$ | $\|\mu\|^2$ | even | odd | $\|\mu_x\|^2$ | $\|\mu_x\|^2_\mathrm{e}$ | $\|\mu_x\|^2_\mathrm{o}$ | $\|\mu_y\|^2$ | $\|\mu_y\|^2_\mathrm{e}$ | $\|\mu_y\|^2_\mathrm{o}$ | $\|\mu_z\|^2$ | $\|\mu_z\|^2_\mathrm{e}$ | $\|\mu_z\|^2_\mathrm{o}$ |
|---|---|---|---|---|---|---|---|---|---|---|---|---|---|---|
| 1 | 17024.0 | -1.00 | 2.33 | 0.00 | 2.33 | 0.00 | 0.00 | 0.0 | 2.33 | 0.0 | 2.33 | 0.00 | 0.00 | 0.0 |
| 2 | 17219.0 | 0.82 | 52.65 | 52.65 | 0.00 | 0.00 | 0.00 | 0.0 | 0.00 | 0.0 | 0.00 | 52.65 | 52.65 | 0.0 |
| 3 | 17222.9 | -0.82 | 5.13 | 5.13 | 0.00 | 0.00 | 0.00 | 0.0 | 0.00 | 0.0 | 0.00 | 5.13 | 5.13 | 0.0 |
| 4 | 17254.2 | -1.00 | 0.05 | 0.05 | 0.00 | 0.00 | 0.00 | 0.0 | 0.00 | 0.0 | 0.00 | 0.05 | 0.05 | 0.0 |
| 5 | 17283.1 | -1.00 | 0.10 | 0.00 | 0.09 | 0.00 | 0.00 | 0.0 | 0.09 | 0.0 | 0.09 | 0.00 | 0.00 | 0.0 |
| 6 | 17448.5 | 1.00 | 1.66 | 1.66 | 0.00 | 0.00 | 0.00 | 0.0 | 0.00 | 0.0 | 0.00 | 1.65 | 1.65 | 0.0 |
| 7 | 17543.7 | 1.00 | 0.26 | 0.26 | 0.00 | 0.26 | 0.26 | 0.0 | 0.00 | 0.0 | 0.00 | 0.00 | 0.00 | 0.0 |
| 8 | 17969.5 | 1.00 | 0.01 | 0.01 | 0.00 | 0.01 | 0.01 | 0.0 | 0.00 | 0.0 | 0.00 | 0.01 | 0.01 | 0.0 |

Without parity mixing only states with $p_k = 1$ should carry oscillator strengths. As can be seen from Tab. 4 this (approximately) holds for the strongest transition $k = 2$ but not in general, indicating parity mixing. Note that this effect is particularly large for the above mentioned resonant states. Since both Coulomb coupling and spectra depend on the transition dipoles we will take a look at the exchange symmetry of the dipoles. Defining even and odd parts of the dipole vectors as

$$\vec{\mu}_{\mathrm{e/o}} = \frac{\vec{\mu}}{2}(1 \pm \mathbf{S}) \tag{S19}$$

and oddness as

$$\nu_\mathrm{o} = \frac{||\mu_\mathrm{o}||}{||\mu||} \tag{S20}$$

gives 0.20 or resolving it via

$$\nu_{\mathrm{o},i} = \frac{||\mu_{\mathrm{o},i}||}{||\mu_i||} \tag{S21}$$

one obtains (0.04,1.00,0.00) for the three vector components, i.e. the dipole has a component that is not symmetric under exchange and this component points along $y$ direction (in accord with Tab. 4).

Note that in principle the exchange parity of Hamiltonian and dipole are not fully equivalent,

i.e. symmetry of the Hamiltonian can be broken not only due to the dipoles but also due to varying positions of the molecules in the two clusters. Rotation around the **c** axis by 180∘ yields perfect matching of atoms of the two clusters for **b** and **c** directions, but a small difference along **a** direction. Overall, the root mean square difference (RMSD) between the geometries of the two clusters is found to be 5.6 Å.

To sum up this discussion and focusing to transitions polarized in $(x, y)$, i.e. $(\mathbf{a}, \mathbf{b})$, directions, we note that (cf. Tab. 4)

- State $k = 1$ has an odd dipole but the parity of that state is odd as well. It has oscillator strength, solely due to the missing exchange parity of the dipole moment between the two clusters; it gives the component in $y$ (**b**) polarization
- State $k = 2$ has an even dipole and parity and carries most oscillator strength. It is $z$ (**c**) polarized, i.e. all dipoles have there largest projection in that direction.
- State $k = 3$ has an even dipole and odd parity. But due to parity mixing (see above) there is some contribution from even parity. Hence the state carries oscillator strength, but less than that of $k = 2$.
- State $k = 7$ is the only $x$ (**a**) polarized transition, it has even parity and an even dipole, with its smallness reflecting the projection of the dipole onto that direction.

## S7.3 Davydov Splitting Analysis

Finally, the results will be put into the context of the Davydov model. It is tempting to consider the two clusters like having an effective dipole and thus forming a Davydov-split pair. So far we have shown that the experimentally observed $x/y$ components can be assigned to the eigenstates with $k = 7/1$. But is this really a Davydov-split pair, i.e. a ± combination of some local cluster states? To investigate this we define local cluster states and project them onto the global eigenstates. The

local cluster states are defined after symmetrization as follows

$$\mathbf{H}_{\mathrm{symm}} = \frac{1}{2}(\mathbf{H}_{11} + \mathbf{H}_{22}) \qquad \mathbf{H}_{\mathrm{symm}}\mathbf{U}_i = E_i^{\mathrm{symm}}\mathbf{U}_i \tag{S22}$$

Next we build Davydov-like ± combinations of alike cluster states as full eigenstates

$$\Phi_{\pm,i} = \frac{1}{\sqrt{2}}\begin{pmatrix} \mathbf{U}_i \\ \pm\mathbf{U}_i \end{pmatrix} \tag{S23}$$

The projection onto full eigenstates $\langle\Psi_j|\Phi_{\pm,i}\rangle$ gives maximum values

$$\begin{aligned} |\langle\Psi_1|\Phi_{-,1}\rangle|^2 &= 0.999 \\ |\langle\Psi_7|\Phi_{+,3}\rangle|^2 &= 0.997 \end{aligned}$$

Apparently, the two transitions do not belong to the same Davydov pair. Instead one finds that

$$\begin{aligned} |\langle\Psi_2|\Phi_{+,1}\rangle|^2 &= 0.912 \\ |\langle\Psi_4|\Phi_{-,3}\rangle|^2 &= 0.946 \end{aligned}$$

That is, the Davydov split pairs are $k = 1, 2$ for $y$-polarization and $k = 4, 7$ for $x$-polarization. Table 5 summarizes the assignment for all Davydov pairs ($w_\pm$ is the squared overlap). Notice that the overlaps are mostly close to 100%, except for states $k = 2, 3$, which is due to the resonance effect mentioned above.

Table 5: Analysis of the eigenstates and transition dipole matrix elements in terms of Davydov-like pair states, Eq. S23. Here $j(\pm)$ denotes that state $\Psi_j$ of the full Hamiltonian which has the largest (squared) overlap $w_\pm$ with the respective state $\Phi_{\pm,i}$

| $i$ | $E_+$ | $E_-$ | $j(+)$ | $j(-)$ | $w_+$ | $w_-$ | $\|\mu\|^2_+$ | $\|\mu\|^2_-$ | $\|\mu_x\|^2_+$ | $\|\mu_y\|^2_+$ | $\|\mu_z\|^2_+$ | $\|\mu_x\|^2_-$ | $\|\mu_y\|^2_-$ | $\|\mu_z\|^2_-$ |
|---|---|---|---|---|---|---|---|---|---|---|---|---|---|---|
| 1 | 17219.0 | 17024.0 | 2 | **1** | 0.91 | 1.00 | 52.65 | 2.33 | 0.00 | 0.00 | 52.65 | 0.00 | **2.33** | 0.00 |
| 2 | 17448.5 | 17283.1 | 6 | 5 | 1.00 | 0.97 | 1.66 | 0.10 | 0.00 | 0.00 | 1.65 | 0.00 | 0.09 | 0.00 |
| 3 | 17543.7 | 17254.2 | **7** | 4 | 1.00 | 0.95 | **0.26** | 0.05 | 0.26 | 0.00 | 0.00 | 0.00 | 0.00 | 0.05 |
| 4 | 17969.5 | 17222.9 | 8 | 3 | 1.00 | 0.89 | 0.01 | 5.13 | 0.01 | 0.00 | 0.01 | 0.00 | 0.00 | 5.13 |

Based on this analysis we conclude that

- indeed the crystal features Davydov split states, but what is observed are not the paired states of the two separate clusters

- $z$-polarization (along $\mathbf{c}$ axis): has the strongest transition for the plus state ($k = 2$), the Davydov pair state is observed in $y$-direction

- $y$-polarization (along $\mathbf{b}$ axis): it is actually the lower (minus) Davydov component ($k = 1$) of the $k = 1, 2$ pair that is observed in the spectrum

- $x$-polarization (along $\mathbf{a}$ axis): here it is the plus state $k = 7$ of the $k = 4, 7$ pair that is visible in the spectrum

- overall the apparent difference between $x$- and $y$-polarization transitions is $E_{+,7} - E_{-,1} =$ 520 cm$^{-1}$, but this energy difference is not related to the coupling strength of a Davydov pair state with respect to the two clusters in the unit cell

# S8 Kinetic Monte Carlo Setup

To study the transport properties of the excitons in the crystal, we need to solve the Master equation which describes the time evolution of the probabilistic processes. One of the most popular methods used for such Markovian processes is the event-driven kinetic Monte Carlo (KMC) method. It is based on sampling of random numbers to set waiting time for the next event and select which event can happen according to the probability proportional to its rate. For each exciton we consider three stochastic events, its intrinsic decay, exciton hopping from one site to another and exciton-exciton annihilation when two excitons occupy the same site. Every event is assigned to a rate constant that reflects its underlying physics. The intrinsic decay rate, i.e. the inverse of the exciton lifetime is taken from the FLIM measurements. Hopping rates are calculated using Förster theory formulated according Fermi's Golden Rule:

$$k_{FT} = \frac{2\pi}{\hbar} V^2 J \tag{S24}$$

where $J$ is the spectral overlap between absorption and emission and $V$ the electronic coupling between donor and acceptor sites.

The spectral overlap $J$ used in Eq. S24 is calculated via[13,14]

$$J = \int_0^\infty \mathcal{A}_{\text{acceptor}}(E)\mathcal{F}_{\text{donor}}(E)\ \mathrm{d}E \qquad , \tag{S25}$$

where $\mathcal{A}$ and $\mathcal{F}$ are the normalized lineshape functions of the acceptor absorption and the donor emission of the states involved in the exciton transfer depending on energy.[15]

Please note that this definition is different from the spectral overlap often used for the Förster transfer in point-dipole approximation of the Coulomb interaction.[16] This is because in Eq. S24 the rate depends on the coupling and not on the interchromophore distance.

To get the lineshape functions $\mathcal{A}(E)$ and $\mathcal{F}(E)$, first, the experimental $\lambda$-dependent spectra $(OD(\lambda), f(\lambda))$ are transferred to energy(E)-dependence, interpolated on an equidistant E-grid

$(OD(E), f(E))$. Then, the lineshape functions are calculated as follows

$$\mathcal{A}(E) \propto \frac{OD(E)}{E} \tag{S26}$$

$$\mathcal{F}(E) \propto \frac{f(E)}{E^5} \tag{S27}$$

With the absorption spectrum of the crystals and their emission spectrum as determined in section S3, the spectral overlap is 1.2/eV. For the monomer it is 0.4/eV in ethanol and 0.5/eV in toluene.

# S9 Further Kinetic Monte Carlo Results

Mean square displacement (MSD) plots and corresponding fits to obtain the diffusion constants are shown for two excitation densities in Fig. S18. The 3D diffusion constant is found to be $3.48 \pm 0.35$ nm$^2$/ps and $3.30 \pm 0.38$ nm$^2$/ps for the higher and lower exciton density, respectively. Both values are close to the average experimental result of 3.80 nm$^2$/ps. In view of the structure of the unit cell, being composed of two clusters with molecular transition dipoles mostly aligned with the **c** crystal axis, one could expect a pronounced anisotropy of the diffusion. In Fig. S19 we show results on MSD and associated diffusion constants for motion along particular crystal directions. In fact the projected diffusion constants are rather similar, i.e. 3.5 (**a**), 2.6 (**b**), and 3.0 nm$^2$/ps (**c**). In other words, diffusion is essentially isotropic.

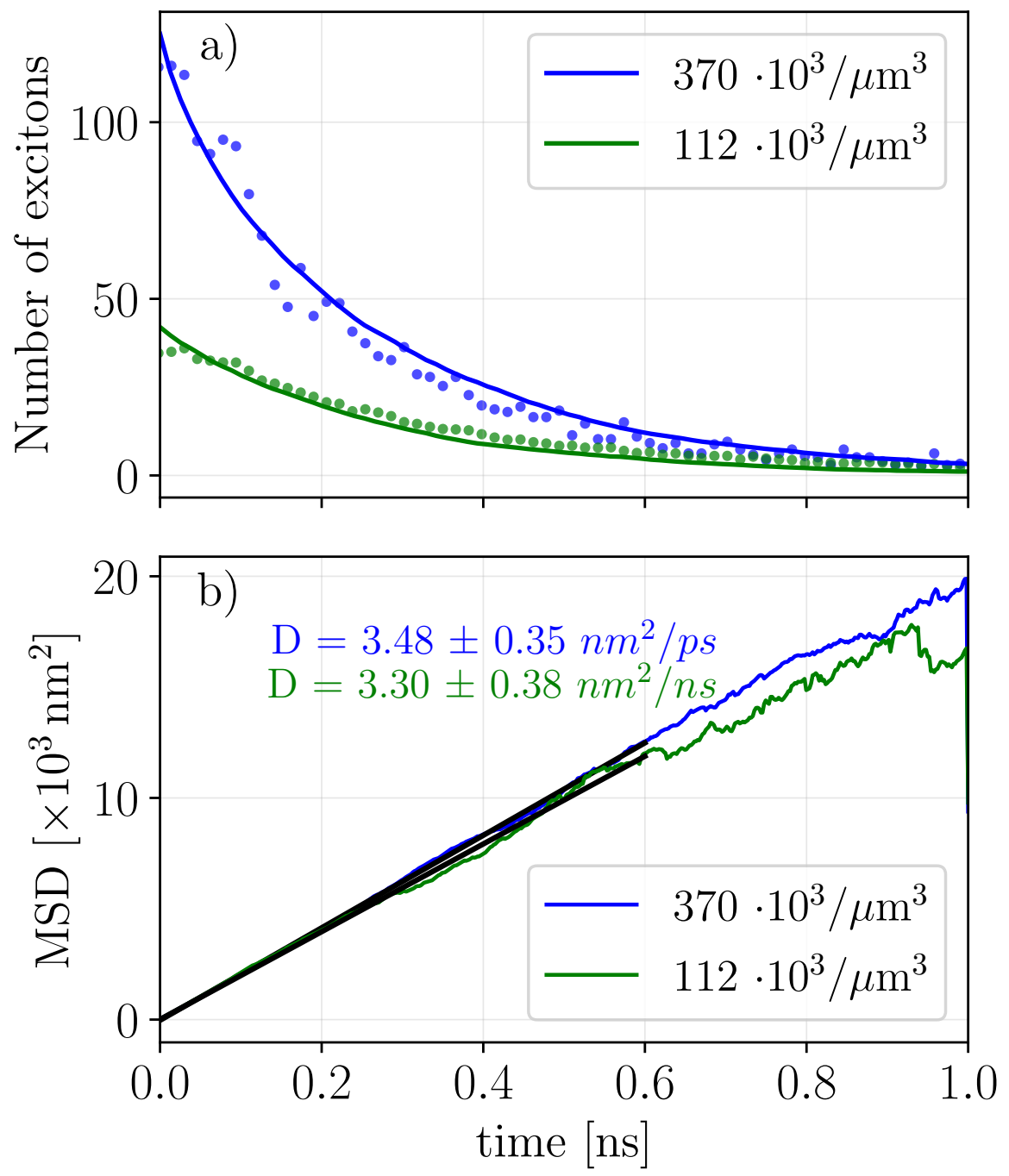


Figure S18: (a) Time evolution of the exciton density for two different initial exciton densities, $370{\cdot}10^3/\mu m^3$ and $112{\cdot}10^3/\mu m^3$. Solid lines represent KMC simulations and dots experimental data. (b) Mean square displacement analysis for these cases. Only the data from 0 to 0.6 ns was considered as more than 80 % of the excitons have already decayed after this time range. The diffusion coefficient is calculated with the Einstein equation MSD$= 6Dt$

.

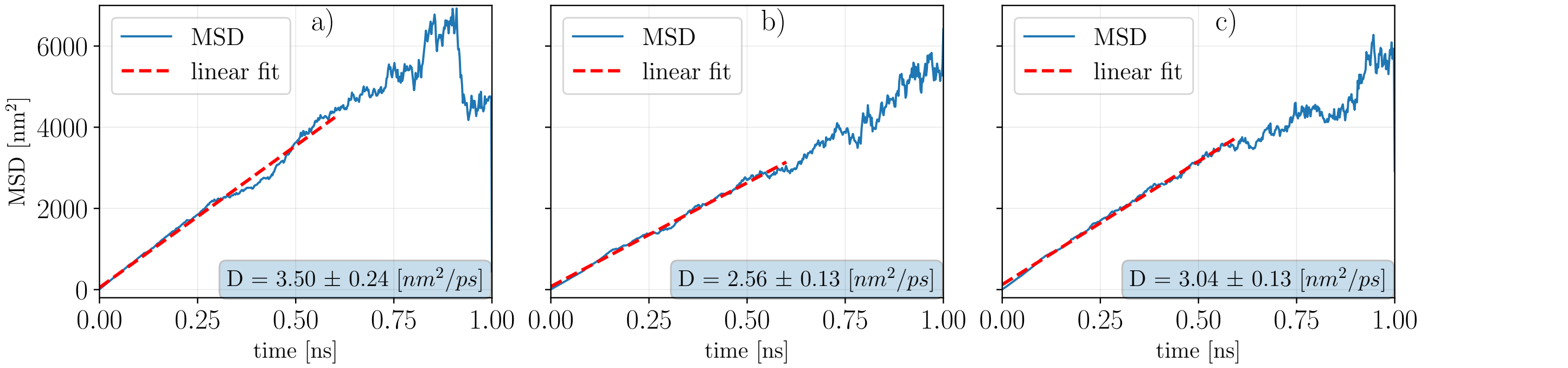


Figure S19: One-dimensional MSD analysis along different crystal axis (**a**, **b**, and **c**, respectively) for the highest exciton density of $370{\times}10^3/\mu m^3$. Only the data from 0 to 0.6 ns is considered since more than 80 % of the excitons have already decayed after this time range.

In order to track the origin of this observation we show the pattern of Förster rates in Fig. 6 of the main text. From that figure we conclude that rates behave essentially isotropic. Closer inspection of possible hopping paths suggests the same order of diffusion constants as seen in Figs. S19.

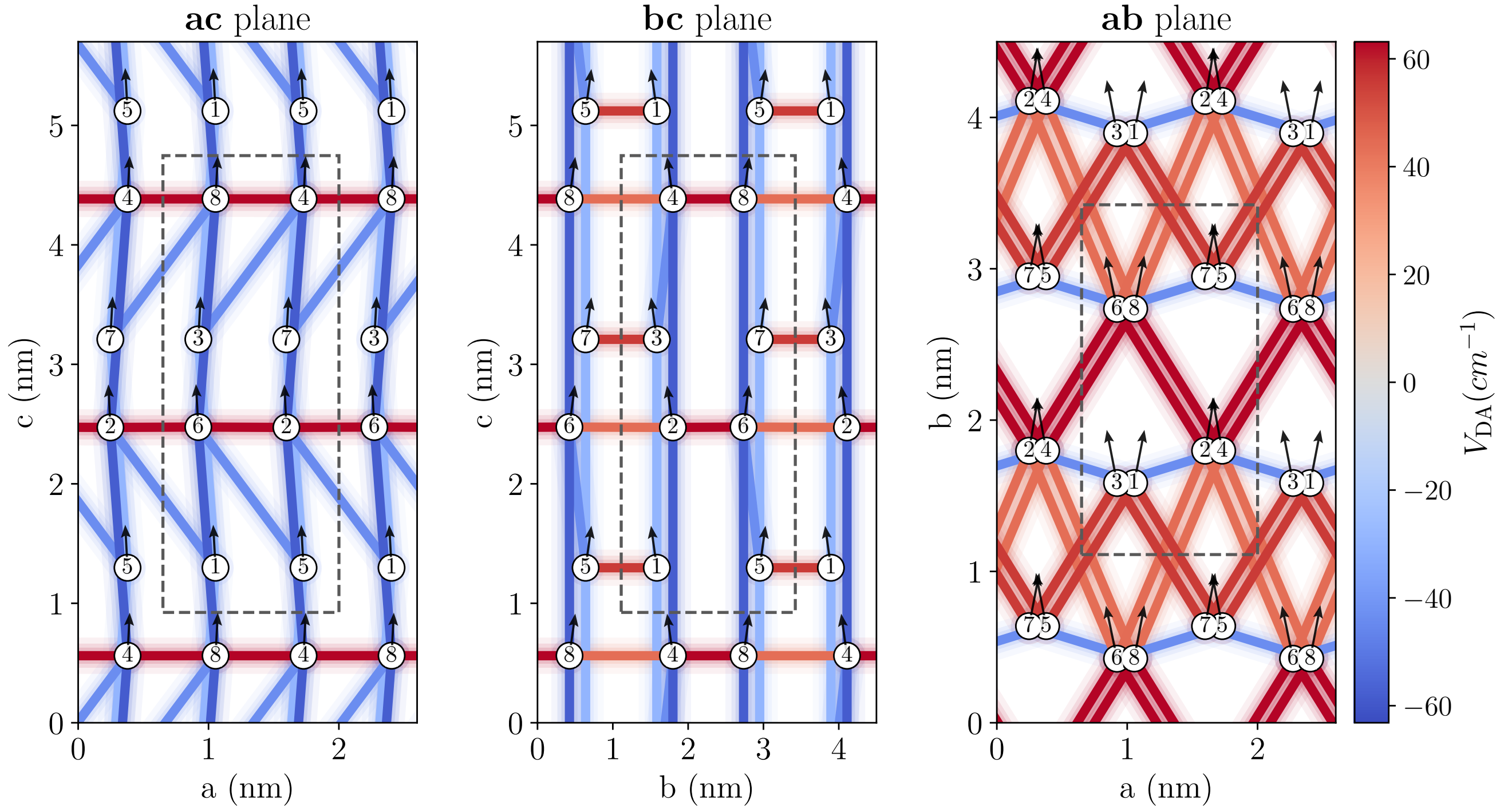


Figure S20: Coulomb couplings $V_{DA}$ between monomers for different crystal planes as indicated. Here 'DA' corresponds to the actual pair of monomers that is connected by the coloured line. The unit cell is shown (dashed rectangle) together with the nearest neighbours.

The pattern of Förster rates, of course, follows that of the squared Coulomb couplings. The (signed) Coulomb couplings are shown in Fig. S20 for different crystal planes. The figure nicely illustrates the $J$- and $H$-type coupling patterns. This figure can be contrasted to the distances between monomers shown in Fig. S21. As already discussed in Fig. S15 the dipole approximation does not hold for the present crystal. Therefore the observation that Coulomb couplings are comparable for different crystal directions, despite the fact that intermonomer distances are rather different, is a nontrivial result of the spatial packing of the monomers in the unit cell.

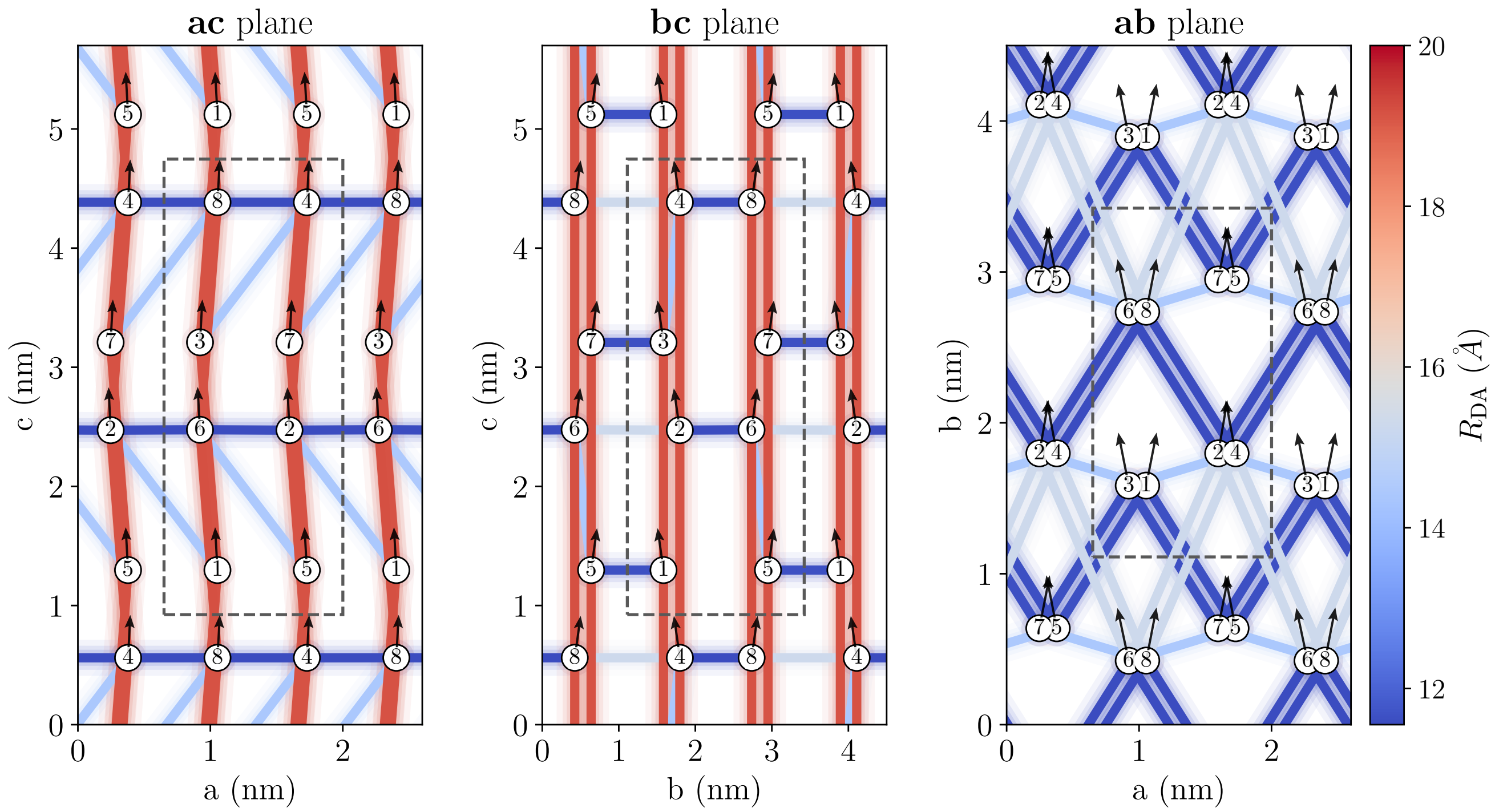


Figure S21: Distances between pairs of monomers for different crystal planes as indicated. Here 'DA' corresponds to the actual pair of monomers that is connected by the coloured line. Only those pairs having sizeable couplings are shown (cf. Fig. S20). The unit cell is shown (dashed rectangle) together with the nearest neighbours.


\end{document}